\documentclass[12pt]{article}
 \pdfoutput=1
\usepackage{amssymb}
\usepackage{amsmath}
\usepackage{amstext}
\usepackage{graphicx,epsfig}
\usepackage{epsfig}
\usepackage{verbatim}
\usepackage{fancybox}
\usepackage{color}
\usepackage{ulem}
\usepackage{enumitem}
\usepackage{subfigure}
\usepackage{bbm}
\usepackage{parskip}
\usepackage{cite}
\usepackage{soul}
\usepackage[utf8]{inputenc}

\newcommand{\Comment}[1]{{}}
\definecolor{MyDarkBlue}{rgb}{0.15,0.15,0.45}
\usepackage[linktocpage=true]{hyperref}
\hypersetup{
colorlinks=true,
citecolor=MyDarkBlue,
linkcolor=MyDarkBlue,
urlcolor=MyDarkBlue,
pdfauthor={},
pdftitle={},
pdfsubject={hep-th}
}

\newcommand{\be}{\begin{equation}}
\newcommand{\ee}{\end{equation}}
\newcommand{\bea}{\begin{eqnarray}}
\newcommand{\eea}{\end{eqnarray}}
\newcommand{\beas}{\begin{eqnarray*}}
\newcommand{\eeas}{\end{eqnarray*}}
\newcommand{\nn}{\nonumber}

\def\({\left(}
\def\){\right)}

\newcommand{\half}{\frac{1}{2}}

\numberwithin{equation}{section}

\def\ba{\begin{eqnarray}}
\def\ea{\end{eqnarray}}

\def\p{\partial}
\def\stu{{St\"uckelberg }}
\def\mpl{M_{\rm Pl}}

\def\nn{\nonumber}

\def\d{\mathrm{d}}

\def\mn{_{\mu \nu}}
\def\ab{_{\alpha \beta}}
\def\mupn{^\mu_{\, \nu}}

\def\L{\mathcal{L}}
\def\stu{St\"uckelberg }

\def\({\left(}
\def\){\right)}
\def\ie{{\it i.e. }}

\def\LPM{\tilde\Lambda_{2}}

\begin{document}

%\maketitle

\begin{center}
{\Large \bf{On the (A)dS Decoupling Limits of Massive Gravity}}\\ \vspace{.2cm}
\end{center}

\vspace{2truecm}

\thispagestyle{empty}
\centerline{{\Large Claudia de Rham,${}^{\rm a,b}$\footnote{\href{mailto:c.de-rham@imperial.ac.uk}{\texttt{c.de-rham@imperial.ac.uk}}} Kurt Hinterbichler,${}^{\rm b,}$\footnote{\href{mailto:kurt.hinterbichler@case.edu}{\texttt{kurt.hinterbichler@case.edu}}} Laura A. Johnson${}^{\rm b,}$\footnote{\href{mailto:lxj154@case.edu}{\texttt{lxj154@case.edu}}}}}
\vspace{.7cm}

 \centerline{{\it ${}^{\rm a}$Theoretical Physics, Blackett Laboratory, Imperial College, London, SW7 2AZ, U.K.}}
 \vspace{.25cm}

 \centerline{{\it ${}^{\rm b}$CERCA, Department of Physics,}}
 \centerline{{\it Case Western Reserve University, 10900 Euclid Ave, Cleveland, OH 44106}}
 \vspace{.35cm}

\begin{abstract}

We consider various decoupling limits of ghost-free massive gravity on (A)dS.
 The first is a decoupling limit on AdS space where the mass goes to zero while the AdS radius is held fixed.  This results in an interacting massive Proca vector theory with a $\Lambda_2\sim (M_{\rm Pl}  m)^{1/2}$ strong coupling scale which is ghost-free by construction and yet can not be put in the form of the generalized Proca theories considered so far.  We comment on the existence of a potential duality between this Proca theory and a CFT on the boundary.  The second decoupling limit we consider is a new  limit on dS, obtained by sending the mass towards the finite partially massless value.  We do this by introducing the scalar St\"uckelberg field which restores the partially massless symmetry.
For generic values of the parameters, only a finite number of operators enter the partially massless decoupling limit and take the form of dS Galileons. If the interactions are chosen to be precisely those of the `candidate' non-linear partially massless theory,
the resulting strong coupling scale has a higher value and the resulting decoupling limit includes an infinite number of interactions which we give in closed form.  These interactions preserve both the linear partially massless symmetry and the dS version of the Galileon shift symmetry.

\end{abstract}

\newpage

\tableofcontents
\newpage

\section{Introduction}
\parskip=5pt
\normalsize
\setcounter{footnote}{0}

Current cosmological observations and the infamous Old Cosmological Constant Problem have motivated the development of many gravitational theories that depart from General Relativity (GR) at large distances. From a field theory perspective, GR is the unique ghost-free theory of an interactive massless spin-2 field in four dimensions, and promoting the graviton to a massive spin-2 field can be seen as a natural possibility. The first consistent model of (soft) massive gravity was proposed almost two decades ago by Dvali, Gabadadze and Porrati (the DGP model \cite{Dvali:2000hr}), a model where the graviton is a resonance. Very rapidly, the DGP model of gravity was extensively explored for its rich cosmological phenomenology as well as a proof of principle of how a spin-2 field could effectively manifest a mass and how the Vainshtein mechanism \cite{Vainshtein:1972sx} could be realized. One of the most powerful developments in DGP was the derivation of its decoupling limit \cite{Luty:2003vm}. Generically, a decoupling limit is a scaling limit that focuses on a given set of interactions at a particular energy scale.  It differs from a low-energy truncation of a theory in that both IR and UV operators may be scaled away.
 The number of degrees of freedom should not increase as one takes the decoupling limit of a theory.  Degrees of freedom may decouple from one another into separate sectors, but they never appear.  When taken properly, a decoupling limit is therefore a powerful tool to constrain the number of degrees of freedom of a theory and study their behavior.
 Within the context of DGP, the decoupling limit was essential in proving the existence of a ghost in the self-accelerating branch of DGP \cite{Koyama:2005tx,Charmousis:2006pn}, understanding the Vainshtein mechanism \cite{Nicolis:2004qq} and providing great insight into the phenomenology of both the normal and self-accelerating branches of DGP. The Friedmann equation in the full DGP model can for instance be derived from  decoupling limit considerations \cite{Chow:2009fm}.

Motivated by the advances pioneered by the DGP model, a multitude of alternative modified gravity models have been proposed more recently. Most of these models can be split into two categories, those for which the graviton is a massless spin-2 field (\ie essentially GR) with additional degrees of freedom that couple to gravity and possibly directly to matter, and those for which the graviton is a massive spin-2 field, which propagates additional polarizations. Among the second class of models, ghost-free massive gravity has emerged as the unique (see Ref.~\cite{deRham:2013tfa}) local and Lorentz-invariant model of gravity in four dimensions where the graviton has a hard mass (as opposed to a resonance) and is ghost free \cite{deRham:2010ik,deRham:2010kj} (see also \cite{Gabadadze:2009ja,deRham:2009rm,deRham:2010gu}, and see \cite{Hinterbichler:2011tt,deRham:2014zqa} for reviews). The emergence of massive gravity was motivated by its decoupling limit which focused on the helicity-0 mode interactions \cite{ArkaniHamed:2002sp,Creminelli:2005qk,deRham:2010ik}. Based on the decoupling limit behavior a full non-linear model was proposed in  \cite{deRham:2010kj} where it was shown to be free of ghosts both in the decoupling limit as well as to all orders in some examples and perturbatively in the general case. The proof of the absence of ghost was then generalized to all orders in the general case in \cite{Hassan:2011hr,Hassan:2011ea}.  As in the case of DGP, the decoupling limit of massive gravity, which was taken on a flat reference metric by sending the graviton mass to zero and the Planck scale to infinity, played an essential role in establishing the consistency and phenomenology of massive gravity \cite{deRham:2010tw,deRham:2012ew}. It was then further extended to a de Sitter reference metric in \cite{deRham:2012kf} which led to the proposal of a unique candidate for partially massless gravity. The full interactions with the vectors were also explored in \cite{Gabadadze:2013ria,Ondo:2013wka}. The decoupling limit of massive gravity on general reference metric was derived in \cite{Fasiello:2013woa} and includes the $\Lambda_3$-decoupling limit on generic cosmological (FLRW) reference metrics as well as on AdS. Within the context of massive gravity or its generalization, the decoupling was also important for cosmology \cite{deRham:2010tw,deRham:2011by,Fasiello:2013woa}, for the study of spherically symmetric solutions \cite{Berezhiani:2013dca} and for radiative stability considerations \cite{deRham:2012ew}.
Motivated by the advances one can reach when analyzing various decoupling limits of a theory we will focus here on two decoupling limits of massive gravity on maximally symmetric spacetimes.

The first decoupling limit we consider is for massive gravity on anti-de Sitter space (AdS). This decoupling limit differs from that derived in \cite{Fasiello:2013woa} where  the AdS curvature scaled as the graviton mass, $L\sim m^{-1}\to 0$. Here we shall take a massless limit $m\rightarrow 0$ while keeping the AdS  radius $L$ held fixed.  In this limit, representation theory of AdS tells us that the five degrees of freedom of the massive spin-2 splits up into a massless spin-2 carrying two degrees of freedom and a massive spin-1 with mass squared equal to $6/L^2$ carrying the remaining three degrees of freedom.   To realize this limit at the level of the Lagrangian, we must introduce a \stu vector field which restores diffeomorphism invariance and becomes the massive vector in the limit. In this limit, we will see the massive spin-1 field fully decouple from the massless spin-2 field and from the conserved stress-tensor. The absence of the vDVZ on AdS is thus manifest, and there is no need to introduce the \stu scalar\footnote{The four \stu fields transform in fact as scalar fields under coordinate transformations. Only in the decoupling limit can the \stu internal global Lorentz invariance  be identified with the spacetime one and the \stu fields split into vectors and scalars under the global Lorentz group.} as is usually done in the flat space decoupling limit.

At the non-linear level, interactions with the tensor are always Planck scale suppressed as compared with vector self-interactions that enter at the scale $\Lambda_2=(\mpl m)^{1/2}$.  Regardless of the form of the potential, the decoupling limit on AdS is thus taken  by sending $\mpl\rightarrow \infty$, $m\rightarrow 0$, with $\Lambda_2$ and the AdS radius held fixed.  In this limit we will obtain a self-interacting massive Proca theory decoupled from the tensors.

For a generic choice of graviton potential or kinetic term this massive Proca theory will be ghostly, propagating four or more degrees of freedom rather than three, reflecting the presence of the Boulware-Deser ghost \cite{Boulware:1973my} (as well as potentially even more problematic ghosts) in the full theory.  But for the Einstein--Hilbert kinetic term and the choice of potential which removes any such ghost \cite{deRham:2010ik,deRham:2010kj}, the resulting massive Proca theory obtained in the AdS decoupling limit gains a non-trivial constraint that removes the extra degree of freedom and leaves only three, the correct number for a massive vector.  This Proca theory contains an infinite set of interactions with all powers of the vector fields, and they are all necessary for the presence of the constraint, which cannot be fully seen at any finite order in powers of the field.  Thus this theory, even though it is ghost free, lies outside of the classification of generalized Proca theories considered thus far \cite{Tasinato:2014eka,Heisenberg:2014rta} and even outside beyond generalized Proca  \cite{Heisenberg:2016eld,Allys:2017map}.

 The other limit we shall be interested in relates to massive gravity on de Sitter space (dS).  On dS, there is a Higuchi bound $m^2\geq 2H^2$ for massive spin-2 \cite{Higuchi:1986py}.  Below the Higuchi bound, $m^2< 2H^2$, the graviton is unstable (except for the massless point $m=0$).   At the Higuchi bound, $m^2=2H^2$, the linear spin-2 field becomes partially massless \cite{Deser:1983tm,Deser:1983mm,Higuchi:1986py,Brink:2000ag,Deser:2001pe,Deser:2001us,Deser:2001wx,Deser:2001xr,Zinoviev:2001dt,Skvortsov:2006at,Skvortsov:2009zu}.  At this point, a scalar gauge symmetry appears which removes one degree of freedom from the linear theory, leaving an exotic irreducible dS representation which propagates four degrees of freedom.  This symmetry does not carry through to the gravitational (non-linear) case \cite{Zinoviev:2006im,deRham:2013wv}, but we will be interested in a partially massless limit $m\rightarrow 2H^2$ in which the dS scale $H$ is held fixed.  In this limit, representation theory of dS tells us that the five degrees of freedom of the massive spin-2 splits up into a partially massless spin-2 carrying four degrees of freedom and a massive scalar with mass squared equal to $-4H^2$ carrying the remaining degree of freedom.   To realize this limit at the level of the Lagrangian, we must introduce a \stu scalar field which restores the partially massless symmetry and becomes the massive scalar in the massless limit.
 
In the full non-linear ghost-free theory of massive gravity, we define $\Delta^2\equiv m^2-2H^2$ to parameterize the difference between the mass and the partially massless mass.  The non-linearities will give operators suppressed by scales made from $\Delta$ and $\mpl$.  The lowest such scale will be $\tilde \Lambda_4=(\mpl \Delta^3)^{1/4}$, which is carried by a cubic scalar self-interaction.
The decoupling limit is then given by taking $\Delta \rightarrow 0$, $\mpl\rightarrow \infty$ with  $\tilde \Lambda_4$ and the dS scale $H$ held fixed.   All that survives is the cubic scalar self-interaction which takes the form of a de Sitter Galileon \cite{Goon:2011uw,Burrage:2011bt,Goon:2011qf}, a generalization of the Galileon \cite{Nicolis:2008in,deRham:2010eu} that lives on de Sitter space yet still possesses a Galileon-like shift symmetry.  The strong coupling scale in this case is $\Lambda=\left(\mpl \Delta^3/H\right)^{1/3}$.

Among the two free parameters, $\alpha_3$, $\alpha_4$ in the dRGT interactions, $\alpha_3$ can be chosen to eliminate to the cubic Galileon at the scale $\tilde \Lambda_4$.  The scale to be fixed is now raised to $\tilde \Lambda_3=(\mpl \Delta^2)^{1/3}$, and the strong coupling scale is $\Lambda=\tilde \Lambda_3$, carried by a quartic dS Galileon.  This interaction can then in turn be eliminated by choosing the other free parameter $\alpha_4$.  This choice of $\alpha_3$, $\alpha_4$ coincides with that of the ``candidate partially massless theory" identified in \cite{deRham:2012kf} and studied in \cite{deRham:2013wv}.  The scale to be held fixed is now $\tilde \Lambda_2=(\mpl \Delta)^{1/2}$ and the strong coupling scale of this theory is $\Lambda=(H \Delta \mpl)^{1/3}$, which is carried by an infinite set of operators, that we provide  in closed form, each involving two powers of the graviton and arbitrary powers of the scalar, as well as the cubic and quartic dS Galileons scalar self-interactions.   All interactions (scalar self-interactions are scalar-tensor interactions) are dS Galileon invariant, and are invariant under the linear partially massless symmetry.  They can be written in terms of the field strength tensor of the partially massless field.  We will also see that they include a spontaneous symmetry breaking potential for the scalar which exhibits a $Z_2$ symmetry breaking which we will discuss.\\

The rest of the manuscript is organized as follows: In section~\ref{linearsecs} we review linear massive gravity on maximally symmetric spacetimes, describe the field content in the different regions of mass--curvature phase plane. We then review the full non-linear theory of massive gravity on (A)dS in section~\ref{sec:nonlinear} before focusing on its AdS decoupling limit in section~\ref{AdSdecsecnonlinear} and partially massless decoupling limit in section~\ref{sec:PM}. A summary of our results is presented in section~\ref{sec:conc}. In Appendix~\ref{app:GFapp} we gather the relevant Lagrangians and tensors that enter the construction of ghost-free massive gravity. We then provide the dS version of the Galileons in appendix~\ref{dsGalileonapps}. Finally in appendix~\ref{adssappendix} we reproduce here for convenience the derivation of the \stu prescription in a maximally symmetric spacetime as provided in \cite{deRham:2012kf}.

\bigskip
{\bf Conventions}:
Throughout this manuscript we work in four spacetime dimensions, and we use the mostly plus metric signature convention, $\eta_{\mu\nu}=(-,+,+,+)$.  Tensors are symmetrized and anti-symmetrized with unit weight, i.e $T_{(\mu\nu)}=\half \left(T_{\mu\nu}+T_{\nu\mu}\right)$,   $T_{[\mu\nu]}=\half \left(T_{\mu\nu}-T_{\nu\mu}\right)$.    On dS space, we denote the dS Hubble scale as $H$, so that $R=12H^2>0$.   On AdS, we denote the AdS radius $L$ so that $R=-{12/ L^2}<0$.  We can go between the two cases with the relation $H^2={-1/L^2}$.

\section{Linear Theory and its Limits\label{linearsecs}}

\begin{figure}[h!]
\begin{center}
\epsfig{file=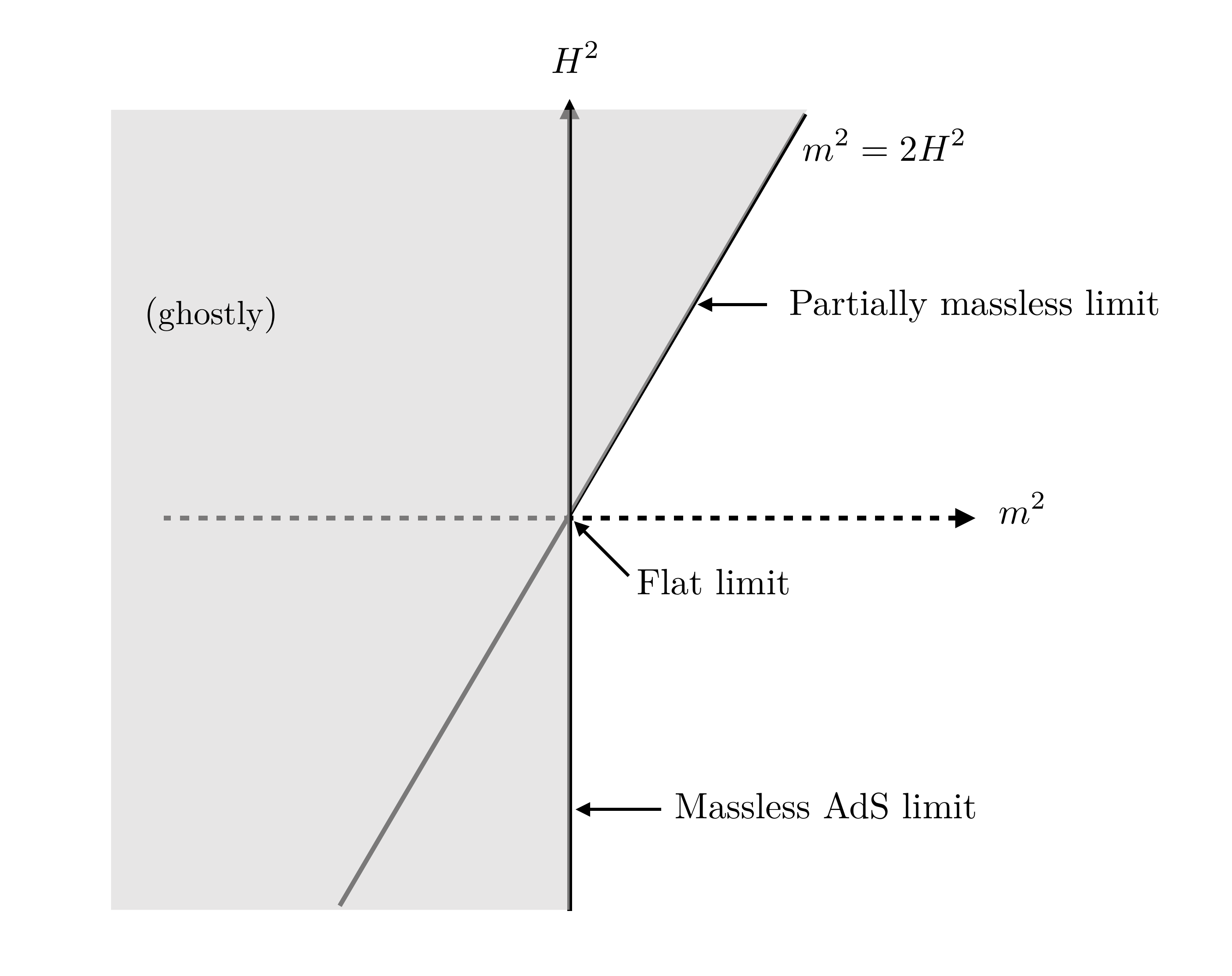,height=4.6in,width=5.9in}
\caption{Linear spin-2 on maximally symmetric spaces and its various limits.}
\label{plot1}
\end{center}
\end{figure}

We start by considering the linearized theory of a massive spin-2 on (A)dS and the various possible decoupling limits.   Decoupling limits occur when we approach lines of enhanced gauge symmetry in the mass vs. background curvature plane (see Fig.~\ref{plot1}).   There are two such lines, the massless line $m^2=0$ and the partially massless line $m^2=2H^2$.  These lines intersect at the origin, and they partition the phase diagram into ghostly and stable regions as shown in Fig.~\ref{plot1}.  There are three different ways of approaching these lines from the ghost-free region, as shown in the figure; we can approach the massless line from the region of negative curvature, we can approach the partially massless line from the region of positive curvature, or we can approach the origin along any angle lying between the two enhanced symmetry lines.

Group theory branching rules will tell us how the five degrees of freedom of the massive spin-2 decompose in each of these limits.  To make manifest the decomposition of the degrees of freedom, we will need to introduce the appropriate \stu fields in order to restore the symmetry which emerges as we approach the enhanced symmetry line.  From the linearized theory we will see how the \stu fields then carry the degrees of freedom removed by the enhanced gauge symmetries, as well as the normalization of their kinetic terms which will set the strong coupling scales.

The Lagrangian for a spin-2 of mass $m$ on (A)dS is given by the Fierz-Pauli Lagrangian \cite{Fierz:1939ix} extended to maximally symmetric curved space \cite{Fang:1978se},
\bea {\cal L}_m&=&  \sqrt{-\gamma} \Bigg[- \frac{1}{2} \nabla_{\alpha} h_{\mu \nu} \nabla^{\alpha} h^{\mu \nu} +\nabla_{\alpha} h_{\mu \nu} \nabla^{\nu} h^{\mu \alpha} - \nabla_{\mu} h \nabla_{\nu} h^{\mu \nu} + \frac{1}{2} \nabla_{\mu} h \nabla^{\mu} h  \nn\\
&&+3H^2(h_{\mu \nu}h^{\mu \nu} -\frac{1}{2} h^2) -\frac{1}{2} m^2(h_{\mu \nu}h^{\mu \nu} - h^2)+{1\over \mpl }h_{\mu\nu}T^{\mu\nu}\Bigg]\, . \label{linearlaggle}
\eea
Here $\gamma\mn$ is the (A)dS background metric with curvature $R\mupn[\gamma]=3H^2\delta\mupn$, while the $h_{\mu\nu}$ is the dynamical spin-2 field. 
Covariant derivatives $\nabla_\mu$ and the raising and lowering of indices is taken with respect to $\gamma\mn$.
We have also added to \eqref{linearlaggle} a coupling to a source $T_{\mu\nu}$ with strength $1/\mpl$.

For generic values of the mass $m$, \eqref{linearlaggle} propagates the 5 degrees of freedom of a fully massive spin-2 on (A)dS space.  However for two special values of $m$ relative to the (A)dS radius, there is (at least within the free linear theory) a reduction in the number of degrees of freedom, giving the enhanced symmetry lines:
\begin{itemize}
\item For $m=0$, there is an enhanced gauge symmetry
\be\label{curvedgaugesy} \delta h_{\mu\nu}=\nabla_\mu\xi_\nu+\nabla_\nu\xi_\mu,\ee
where $\xi_{\mu}$ is a vector gauge parameter.  The action propagates 2 degrees of freedom of a massless spin-2 on (A)dS.   This is the nothing but general relativity with a cosmological constant linearized about its maximally symmetric solution, and the symmetry \eqref{curvedgaugesy} is linearized diffeomorphism symmetry.

At this massless line, consistency demands that the source is conserved,
\ba
\label{eq:DTm}
 \nabla_\nu T^{\mu \nu}=0\, .
\ea

\item For $m^2=2H^2$ there is an enhanced gauge symmetry
\be
\label{partialmasslesssym}
\delta_\chi h_{\mu\nu}=\nabla_\mu\nabla_\nu\chi+{H^2}\chi \gamma_{\mu\nu},
\ee
where $\chi$ is a scalar gauge parameter.  The theory at $m^2=2H^2$ is called {partially massless}.  The gauge symmetry \eqref{partialmasslesssym} kills one degree of freedom, so the partially massless graviton carries 4 degrees of freedom rather than 5.

At this partially massless line, consistency demands that the source satisfies
\ba
\label{eq:DTPM}
\nabla_\mu \nabla_\nu T^{\mu \nu}=-H^2 T^\mu{}_\mu\,.
\ea
(In the limit where $H\to 0$, this is consistent with conservation \eqref{eq:DTm}.)
\end{itemize}

Away from the enhanced symmetry lines the source need not necessarily satisfy any conservation conditions.  However, as we will see, we will generally find strong coupling in the various limits unless we demand weaker conditions such as $\nabla_\mu T^{\mu \nu}=\mathcal{O}(m^2)$ that are consistent with the required conservation in the various limits.

The theory \eqref{linearlaggle} has ghost-like instabilities when $m^2<0$, and is stable when $m=0$.  For $m^2>0$, there are two different cases: on AdS and flat space ($H^2\leq 0$), the theory is stable for $m^2>0$.  On dS ($H^2>0$), the theory is unstable for $0<m^2<2H^2$ and stable for $m^2\geq 2H^2$.  The massless and partially massless lines form boundaries for the stable regions, as shown in Fig.~\ref{plot1}.

Looking at the lines of enhanced gauge symmetry in Fig.~\ref{plot1}, we can consider three different limits in which we approach these lines from the stable region; a massless limit from AdS, a partially massless limit from dS, and a massless limit along some direction from the stable region.   In these limits, the massive graviton will be breaking up into the (partially) massless graviton and the remaining degrees of freedom.  We will now consider these limits in turn.

\subsection{Flat space massless limit}

The flat space massless limit is the usual decoupling limit considered in studies of massive gravity.  In the flat space massless limit, a massive spin-2 breaks up into a massless spin-2, a massless spin-1, and a massless scalar.  Thus we expect to need two different \stu fields; one to carry the massless spin-1, and one to carry the massless spin-0.

Because the massless, flat space point is co-dimension two in the mass vs. curvature plane of Fig.~\ref{plot1}, we can approach this point in various directions: $m\rightarrow 0,\ H\rightarrow 0$ keeping the ratio $H/m$ 
fixed but arbitrary.   The limit will be direction dependent.

Since we recover linearized diffeomorphisms at $m=0$, the best way to diagnose what occurs in the $m\to 0$ limit is to introduce the \stu field $A_\mu$ that restores this local symmetry, as well as a \stu field $\phi$ which restores the $U(1)$ necessary for the expected massless spin-1 mode,
\be
h_{\mu\nu}\rightarrow h_{\mu\nu}+\frac{1}{m}\(\nabla_\mu A_\nu+\nabla_\nu A_\mu\)\, ,\ \ \ A_\mu\rightarrow A_\mu+\frac 1m \nabla_\mu\phi\,. \label{flatsturepe}
\ee
The mass scaling here is chosen so that the kinetic terms for the fields come out canonically normalized.

After the replacement \eqref{flatsturepe} there is a gauge symmetry (even away from $m=0$),
\bea \delta h_{\mu\nu}=\nabla_\mu \xi_\nu+\nabla_\nu \xi_\mu,\ \
\delta A_\mu=\nabla_\mu\zeta-m\xi_\mu ,\ \ \ \delta\phi=-m\zeta, \label{massivenolimitstuk}
\eea
for the gauge parameters $\xi_\mu$ and $\zeta$.

The flat space massless decoupling limit is taken by sending
\be m\rightarrow 0,\ \ \ H\rightarrow 0,\ \ \ {m\over H}\ {\rm fixed},\ \ \  h\mn,\ A_\mu,\ \phi \ \ {\rm fixed}.\label{masslessdeclimlee}
\ee
In this limit, the resulting  Lagrangian, after a conformal transformation to diagonalize the kinetic terms,
\be h_{\mu\nu}\rightarrow h_{\mu\nu}+\phi \, \eta_{\mu\nu}\, , \label{conftr}\ee
is the flat space Lagrangian for the expected massless spin-2, massless spin-1, and massless scalar,
\bea   {\cal L}= {\cal L}_{H,m=0} -{1\over 2} F_{\mu\nu}^2   -3\left(1-2\frac{H^2}{m^2}\right) (\partial\phi)^2+\frac{1}{\mpl}h_{\mu\nu}T^{\mu\nu}+{1\over 2\mpl} \phi T \, ,\label{masslesslimeqscnoce}
\eea
where $F_{\mu\nu}\equiv\partial_\mu A_\nu-\partial_\nu A_\mu$ is the standard Maxwell field strength for the massless spin-1.

We have assumed that the divergence of the stress tensor vanishes faster\footnote{These requirements are equivalent to requiring that the matter coupling preserve the appropriate symmetries in those limit.} than ${\cal O}(m)$ in the limit, and its double divergence faster than ${\cal O}(m^2)$, otherwise we would get strong coupling between the vector and the divergence of the stress tensor, or between the scalar and the double divergence.   The coupling between the scalar and the trace of the stress tensor comes from the necessary conformal transformation \eqref{conftr} and is responsible for the vDVZ discontinuity \cite{vanDam:1970vg,zakharov}.  Note that the scalar kinetic term in \eqref{masslesslimeqscnoce} depends on $m/H$ and hence on the direction of the limit.  Once the scalar is canonically normalized, the coupling will go like $\sim {1\over 2\mpl\sqrt{1-2\frac{H^2}{m^2} }} \phi T $, and so
 we interpolate between having a vDVZ discontinuity due to the $\phi T$ coupling when $H\rightarrow 0$ first, no vDVZ discontinuity\footnote{Notice however that sending $m\to 0$ faster than $H\to 0$  is only a viable limit on AdS where $H^2<0$ and not on dS.} when $m\rightarrow 0$ first \cite{Higuchi:1986py,Porrati:2000cp,Kogan:2000uy,Karch:2001jb}, and strong coupling at the partially massless point $m^2=2H^2$ unless the trace scales to zero faster than ${\cal O}(m)$.  This point will be discussed further later.

Note that the transformation for $\phi$ in \eqref{massivenolimitstuk} goes to zero in the limit \eqref{masslessdeclimlee}, and hence $\phi$ becomes gauge invariant, whereas the tensor $h_{\mu\nu}$ and vector $A_\mu$ maintain the standard flat space massless symmetry transformations,
\bea \delta h_{\mu\nu}=\partial_\mu \xi_\nu+\partial_\nu \xi_\mu,\ \
\delta A_\mu=\partial_\mu\zeta ,\ \ \ \delta\phi=0\,. \label{massivenolimitstukmassless3}
\eea

\subsection{AdS massless  limit\label{linadssec}}

In this limit, we take the graviton mass to go to zero, $m\rightarrow 0$, while leaving the AdS radius $L^2=-1/H^2$ fixed.

Group theoretically, the massive spin-2 representation decomposes in this limit into a massless spin-2 and a massive vector of mass squared $6/L^2$.  Representations in AdS \cite{doi:10.1063/1.1705183} can be labelled as $(\Delta_c,s)$, where $s$ is the spin and $\Delta_c$ is the dimension of the dual CFT operator, in terms of which the mass is given by $m^2L^2=(\Delta_c +s-2)(\Delta_c -s-1)$ for $s\geq 1$ and $m^2L^2=\Delta_c (\Delta_c -3)$ for $s=0$. The limit we are interested in corresponds to the branching rule \cite{doi:10.1063/1.1705183,Fronsdal:1978vb}
\be (\Delta_c ,2)\underset{\Delta_c \rightarrow 3}{\longrightarrow} (3,2)\oplus (4,1)\, .\label{gtdecasdde}\ee
Here $(\Delta_c ,2)$ can be thought of as a non-conserved symmetric traceless primary operator $T_{ij}$ in the dual three dimensional CFT.  As it approaches its conserved dimension $\Delta_c =3$, the divergence $\partial ^j T_{ij}$, which has dimension $\Delta_c +1$, decouples and becomes its own primary, becoming the $(4,1)$.

Based on this group theoretical decomposition, we expect to only need to introduce a vector \stu $A_\mu$, patterned after the gauge symmetry \eqref{curvedgaugesy} which is restored in the $m=0$ limit,
\be h_{\mu\nu}\rightarrow h_{\mu\nu}+\frac{1}{m}\(\nabla_\mu A_\nu+\nabla_\nu A_\mu\)\,. \label{massiveveddcle}
\ee
We do not need to introduce the scalar with the associated $U(1)$ symmetry as we did in \eqref{flatsturepe} for the massless case, since the vector is massive in the $m\rightarrow 0$ limit and the scalar degree of freedom does not appear.   The mass scaling here is chosen so that the kinetic term for the vector comes out canonically normalized.

After \eqref{massiveveddcle}, the theory has a gauge symmetry
\be
\delta h_{\mu\nu}=\nabla_\mu\xi_\nu+\nabla_\nu\xi_\mu,\ \ \ \delta A_{\mu}=-m\xi_\mu\,. \label{linermadssymde}
\ee
We then take the limit
\be m\rightarrow 0,\ \ \ L\ \ {\rm fixed},\ \  h\mn,\ A_\mu \ \ {\rm fixed}\,, \label{mmlinelime}
\ee
and the Lagrangian in this limit becomes 
\be
{\cal L}= {\cal L}_{m=0}+ \sqrt{-\gamma}\left[ -{1\over 2} F_{\mu\nu}^2 - {6\over L^2} A^2+{1\over \mpl }h_{\mu\nu}T^{\mu\nu}\right]\, .
\ee
We see explicitly the appearance of a massless graviton and a massive vector with the mass
\be  m_A^2={6\over L^2}.\ee
The decoupling limit is smooth without the need to introduce the scalar, as expected from the branching rule \eqref{gtdecasdde}; before the limit there are 5 degrees in the massive graviton, after the limit there are 2 degrees of freedom in the massless graviton and 3 degrees of freedom in the massive vector.

We have assumed that the divergence of the stress tensor vanishes faster than ${\cal O}(m)$ in the limit, otherwise we would get strong coupling between the vector and the divergence of the stress tensor.
There is no coupling between the \stu degrees of freedom and the source, so as is well-known, there is no vDVZ discontinuity on AdS \cite{Higuchi:1986py,Porrati:2000cp,Kogan:2000uy,Karch:2001jb,deRham:2016plk} (see however \cite{Dilkes:2001av,Duff:2001zz} for studies of one-loop corrections to this statement).

Note that the transformation for $A_\mu$ in \eqref{linermadssymde} goes to zero in the limit \eqref{mmlinelime}, and hence becomes gauge invariant, whereas the tensor $h_{\mu\nu}$ maintains the linear massless symmetry,
\be
\delta h_{\mu\nu}=\nabla_\mu\xi_\nu+\nabla_\nu\xi_\mu,\ \ \ \delta A_{\mu}=0\,. \label{linermadssymdelimit}
\ee

\subsection{Partially massless limit\label{PMlimitsec}}

In this limit, we let the graviton mass approach its partially massless value $m^2\rightarrow m_{\rm PM}^2=2H^2$, while leaving the de Sitter scale $H$ fixed.

Group theoretically, the massive spin-2 representation decomposes in this limit into a partially massless spin-2 and a massive scalar of mass squared $-4H^2$.  Using the AdS representation notation referred to at the beginning of Section \ref{linadssec}, the branching rule is
\be (\Delta_c ,2)\underset{\Delta_c \rightarrow 2}{\longrightarrow} (2,2)\oplus (4,0)\, .\label{gtdecasddepm}\ee
The value $\Delta_c =2$ for a spin-2 primary $T_{ij}$ is the value at which it satisfies the double conservation condition $\partial_i\partial_j T^{ij}=0$, and is dual to a partially massless graviton\footnote{Higher spin AdS/CFT examples including partially massless states have been studied in \cite{Bekaert:2013zya,Alkalaev:2014nsa,Basile:2014wua,Joung:2015jza,Brust:2016gjy,Brust:2016zns,Brust:2016xif,Basile:2018acb}.} \cite{Dolan:2001ih}.  Here $(\Delta_c ,2)$ is a non-conserved stress tensor $T_{ij}$ in the dual three dimensional CFT, and as it approaches its double conserved dimension $\Delta_c =2$, the double divergence $\partial ^i \partial^jT_{ij}$, which has dimension $\Delta_c +2$, decouples and becomes its own scalar primary, becoming the $(4,0)$.

Based on this group theoretic decomposition, to restore the partially massless symmetry we expect to only need to introduce a \stu scalar, and no vector.
Since the linear theory recovers the partially massless gauge symmetry \eqref{partialmasslesssym} at the partially massless point, we introduce the scalar \stu field patterned after that symmetry,
\be
h_{\mu\nu}\rightarrow h_{\mu\nu}+{1\over H\Delta}\left( \nabla_\mu \nabla_\nu \phi+H^2 \gamma_{\mu\nu} \phi\right)\,. \label{PMlinestue}
\ee
Here we have introduced the quantity
\be \Delta^2=m^2-2H^2\, ,
\ee
so that $\Delta=0$ is the partially massless value.  The prefactor in front of $\phi$ in \eqref{PMlinestue} is chosen so that $\phi$ will come out with canonical normalization.

The theory \eqref{PMlinestue} now has a gauge symmetry
\be \delta h_{\mu\nu}=\nabla_\mu\nabla_\nu\chi+{H^2}\chi \gamma_{\mu\nu},\ \ \ \delta  \phi=-H\Delta\,\chi. \label{PMstymsrese}
\ee
We now take the partially massless limit
\be
\Delta\rightarrow 0,\ \ \ H \ \ {\rm fixed},\ \ h\mn, \  \phi \, \  {\rm fixed}\, . \label{pmdeclimlineare}
\ee
The Lagrangian in this limit becomes
\bea   {1\over \sqrt{-\gamma}}{\cal L}&=& {1\over \sqrt{-\gamma}}{\cal L}_{m_{\rm PM}}+h_{\mu\nu}T^{\mu\nu}\nn\\
&& + 3\left(-{1\over 2}(\nabla \phi)^2+2H^2 \phi^2\right)+ \,  {1\over H\Delta}\phi \(\nabla_\mu \nabla_\nu T^{\mu\nu}+H^2 T\)\, . \label{linearpmlimitdece}
\eea
We see the appearance of a partially massless graviton and a scalar with mass
\be  m_\phi^2=-4H^2\, ,\ee
as expected from the branching rule \eqref{gtdecasddepm}.
The decoupling limit is smooth; before the limit there are 5 degrees of freedom in the massive graviton, after the limit there are 4 degrees of freedom in the partially massless graviton and 1 degree of freedom in the massive scalar.  As mentioned, there is no need to introduce a vector \stu field in this case, and the decoupling limit is already smooth without it, as expected from the group theoretic decomposition \eqref{gtdecasddepm}.

Note that the coupling to $T^{\mu\nu}$ in the second line of \eqref{linearpmlimitdece} blows up, and we have strong coupling in the decoupling limit, unless the stress-energy satisfies the modified conservation relation \eqref{eq:DTPM} on de Sitter to ${\cal O}\left(\Delta\right)$ in the limit. We recall that while the partially massless line is indeed a line of enhanced symmetry, the symmetry recovered at the linear level in that limit is the partially massless one rather than diffeomorphism invariance. As a result there is no reason to have the stress-energy tensor being conserved on the partially massless line, rather we would like its coupling to the spin-2 field to preserve the partially massless symmetry which is precisely what  the modified conservation relation \eqref{eq:DTPM}  dictates.
In taking the partially massless limit, we should thus ensure that the stress tensor satisfies this requirement.

Finally, note that the transformation for $\phi$ in \eqref{PMstymsrese} goes to zero in the limit \eqref{pmdeclimlineare}, and hence becomes gauge invariant, whereas the tensor $h_{\mu\nu}$ maintains the linear partially massless symmetry,
\be \delta h_{\mu\nu}=\nabla_\mu\nabla_\nu\chi+{H^2}\chi \gamma_{\mu\nu},\ \ \ \delta  \phi=0. \label{PMstymsresel}
\ee

\section{Non-Linear Massive Gravity on (A)dS}
\label{sec:nonlinear}

The previous considerations were at the level of free (linear) theories with coupling to an external non-dynamical stress-tensor. As is well-known, as soon as the degrees of freedom that enter the stress-tensor are made dynamical, i.e. as soon as the spin-2 field interacts, we are led to a gravitational theory with an infinite number of interactions of a precise form.
We now turn to the fully interacting theory of massive gravity on (A)dS, and the fully non-linear versions of the various limits described in section \ref{linearsecs}.

The only known ghost-free interacting effective field theory of a single massive spin-2 in four dimensions is ghost-free massive gravity \cite{deRham:2010kj}, also known as dRGT theory (see \cite{Hinterbichler:2011tt,deRham:2014zqa} for reviews).  Placing it on an (A)dS background \cite{Hassan:2011tf},  the Lagrangian is
\ba {\cal L}={\mpl ^2\over 2}\sqrt{-g}\left( R[g]-6H^2 +{m^2}\left[S_2({\cal K})+\alpha_3 S_3({\cal K})+\alpha_4 S_4({\cal K})\right]\right)+
\L_{\rm matter}(g\mn, \psi_i). \label{nonlinlag}
\ea
Here $\psi_i$ denote any matter fields, and their corresponding stress-energy tensor is covariantly conserved on the matter equations of motion, 
though we will not be explicitly concerned with the matter in what follows.  The tensor ${\cal K}\mupn$ is given by
\be {\cal K}^\mu_{\ \nu}=\delta^\mu_{\ \nu}-\left(\sqrt{g^{-1} \gamma}\right)^\mu_{\ \nu}\, ,\ee
with $g_{\mu\nu}$ the dynamical metric and $\gamma_{\mu\nu}$ the background (A)dS metric, whose constant curvature scale is $H$, $R[\gamma]=12H^2$.   The $S$'s are the standard symmetric polynomials, defined in Appendix~\ref{app:GFapp}.
There are two dimensionless parameters, $\alpha_3$ and $\alpha_4$, governing the interactions of the theory.   Expanding $g_{\mu\nu}=\gamma_{\mu\nu}+{2\over \mpl}h_{\mu\nu}$, \eqref{nonlinlag} reduces to \eqref{linearlaggle} at quadratic order in $h_{\mu\nu}$.

We will now proceed to investigate the fully non-linear versions of the various decoupling limits described in section \ref{linearsecs}.  In all the cases, the non-linear decoupling limit comes with taking $\mpl \rightarrow \infty$ while leaving some intermediate scale, the strong coupling scale, fixed. This strong coupling scale is always a combination of the Planck scale, the mass of the spin-2 field and the background curvature scale and is kept finite in the  $\mpl \rightarrow \infty$  limit by simultaneously sending either the mass or a combination of the mass and the curvature to zero.

The flat-space massless decoupling limits have already been worked out elsewhere; the case $H/m\to 0$ is part of the original studies of ghost-free massive gravity \cite{deRham:2010ik,deRham:2010kj} and the $H/m\not\to0$ generalization was derived in \cite{deRham:2012kf}. In what follows we start with the AdS decoupling limit in Section \ref{AdSdecsecnonlinear}, before moving on to the partially massless decoupling limit in section \ref{sec:PM}.

\section{Non-Linear Massless AdS Decoupling Limits\label{AdSdecsecnonlinear}}

We start with the massless decoupling limit of the full non-linear theory on AdS.  This requires the introduction of the fully non-linear \stu fields, and will result in an interacting massive Proca theory in the decoupling limit.

\subsection{\stu in AdS}

On an AdS background, the \stu fields take a more complicated form than in flat space.  The derivation of their appropriate form in AdS follows the prescription described in \cite{deRham:2012kf}, which we review in AdS in Appendix \ref{adssappendix}. The end result is that the \stu field $A_\mu$ is to be introduced through the reference metric $\gamma_{\mu\nu}$ via the replacement
\ba
\gamma_{\mu\nu}\rightarrow \tilde{\gamma}_{\mu \nu}={\gamma}_{\mu \nu}-S_{\mu\nu}-S_{\nu\mu}+S_{\mu}^{\ \lambda}S_{\nu\lambda}-{1\over L^2+ \bar A^2}T_{\mu} T_{\nu}, \label{nonlineadsstue}
\ea
with
\ba
S_{\mu\nu}=  \nabla_\mu  \bar A_\nu+\gamma_{\mu\nu}\left(1-\sqrt{1+{1\over L^2} \bar A^2}\right),\quad
T_{\mu}= {1\over 2}\partial_\mu (\bar A^2)-\sqrt{1+{1\over L^2} \bar A^2}\  \bar A_\mu\,,  \label{nonlineadsstue2}
\ea
where the covariant derivatives and raised/lowered indices are with respect to the reference metric $\gamma_{\mu\nu}$, and
\be \bar A_\mu= {2\over \mpl m }A_\mu\, , 
\ee
results in a canonically normalized $A_\mu$. In the decoupling limit $A_\mu$ will play the role of a vector field, but intrinsically it is related to the \stu scalar fields.

Making this substitution in the non-linear ghost-free massive gravity action \eqref{nonlinlag} with
${\cal K}^\mu_{\ \nu}=\delta^\mu_{\ \nu}-\left(\sqrt{g^{-1} \tilde\gamma}\right)^\mu_{\ \nu}$
 and expanding the metric about the AdS vacuum solution
\be g_{\mu\nu}=\gamma_{\mu\nu}+{2\over \mpl}h_{\mu\nu}\, ,
\ee 
results in a vector-tensor theory.  By construction, the \stu field restores full non-linear diffeomorphism invariance and hence does not appear through the Einstein-Hilbert plus cosmological constant part of the action \eqref{nonlinlag}, and so the only contributions come through the non-diff-invariant mass terms in \eqref{nonlinlag}.   As we learned from the linear theory in Section \ref{linadssec}, there is no need to further introduce a longitudinal mode.

\subsection{Interacting Proca in the decoupling limit}

The interactions among the tensors and vectors that appear after the \stu replacement are schematically of the form
\be {\cal L}_{l,n}\sim  m^2 \mpl^2 \(\partial+{1\over L}\)^{n} {1\over \mpl^l }{1\over (\mpl m)^n}  h^l A^n \,, \ \ l,n\geq 0,\ \ \ l+n\geq 3\,. \ee
(Here we treat $\partial$, $1/L$ as schematically identical, since they are held fixed in the limit of interest and can be converted into each other by commuting derivatives.)
The strong coupling scale is the lowest such scale that appears.  As is well known, this strong coupling scale is \cite{deRham:2016plk,deRham:2015ijs,Gabadadze:2017jom}
\be
\Lambda_2 =\left(\mpl m\right)^{1/2}.
\ee
 The operators in the expansion that carry this scale are the vector self-interactions, 
\be
{\cal L}_{0,n} \sim {1\over \Lambda_2^{2n-4}} \(\partial+{1\over L}\)^{n}   A^n\, ,\ \ \ n\geq 3 \,. \label{vecselfintsce}
\ee

The $\Lambda_2$ decoupling limit is then
\be \mpl \rightarrow \infty,\ \ \ m\rightarrow 0,\ \ \ L,\ \Lambda_2\ \ {\rm fixed},\ \ \   h\mn,\  A_\mu \  {\rm fixed.}\ee
In this limit, all interactions that involve a tensor vanish, since they are suppressed by further powers of the Planck scale, and what is left is an infinite number of vector self-interactions\footnote{There are also an infinite number of terms involving the vector mode in the decoupling limit of massive gravity on flat spacetime \cite{Gabadadze:2013ria,Ondo:2013wka}, though in this case they are vector-scalar interactions rather than vector self-interactions.} of the form \eqref{vecselfintsce}.  Note that since the stress-energy tensor for a standard diffeomorphism invariant matter coupling is covariantly conserved, so the Proca field does not interact with matter in this decoupling limit. In this strict limit, we are therefore dealing with an isolated self-interacting Proca field theory living on AdS without any outside interaction, either to gravity or to standard matter.

This decoupling limit Proca Lagrangian is
\be {\cal L}_{\Lambda_2}=\mathcal{L}^{(2)}_{\Lambda_2}( A)+{1\over \Lambda_2^2 }\mathcal{L}^{(3)}_{\Lambda_2}( A)+{1\over \Lambda_2^4}\mathcal{L}^{(4)}_{\Lambda_2}( A)+\cdots,\label{fullProcaeseq}\ee
Where the terms up to quartic order in the fields are given by 
\ba
{1\over \sqrt{-\gamma}}\mathcal{L}^{(2)}_{\Lambda_2}(A) &=&  -{1\over 2} F_{\mu\nu}^2 - {6\over L^2} A^2\, , \label{linearzedlagliste}\\
{1\over \sqrt{-\gamma}}\mathcal{L}^{(3)}_{\Lambda_2}(A) &=&
\frac{\alpha_3}{2}S_3(B)
-\frac {1}{2} F^{\mu \alpha}F^\nu{}_\alpha X^{(1)}\mn(B)
-\frac{3}{L^2}A^2 B
\,,\label{eq:Proca3p}
\\
{1\over \sqrt{-\gamma}}\mathcal{L}^{(4)}_{\Lambda_2}(A)   &=&
\frac{1}{8}\((F_{\mu\nu}F^{\mu\nu})^2-F_{\mu\nu}F^{\nu\alpha}F_{\alpha\beta}F^{\beta\mu}\)+\frac{\alpha_4}{2} S_4(B)-\frac{3\alpha_3}{4}F^{\mu\alpha}F^\nu{}_\alpha X^{(2)}\mn(B)\nn   \\  \nn
 &+&   \frac{1}{4}F^{\mu\alpha}F^{\nu\beta}B\mn B\ab +\frac1{4}F^{\mu\alpha}F^{\nu}{}_{\alpha}B^2\mn-\frac{1}{2} F^{\mu\alpha}F^{\nu\alpha}B B\mn \\ \nn
&-&  \frac{1+6\alpha_3}{L^2}A^2 S_2(B)+\frac{1+3\alpha_3}{L^2}A^\mu A^\nu X^{(2)}\mn(B) \\
&+&   \frac{2}{L^2}\(A^2 F_{\mu\nu}F^{\mu\nu}-A^\mu A^\nu B_{\mu\alpha}F_\nu{}^\alpha+\frac 12 A^\mu A^\nu F_{\mu\alpha}F_\nu{}^\alpha\)+\frac{12}{L^4}A^4\,.  \label{eq:Proca4}
\ea
Here we have defined $F\mn=\nabla_{\mu}A_\nu-\nabla_{\nu}A_\mu$ and $B\mn=\nabla_{\mu}A_\nu+\nabla_{\nu}A_\mu$, with $B=B^\alpha{}_\alpha$, and the tensors $X^{(n)}\mn$ and the characteristic polynomials $S_n$ are defined in appendix \ref{app:GFapp}.

Note that the scale suppressing the interactions would be $\Lambda_2$ even for a generic gravitational potential, i.e. a potential not of the ghost-free dRGT form.  The decoupling limit Lagrangian would still be a self-interacting Proca theory with terms of the form \eqref{vecselfintsce}.  This is in contrast to the flat space decoupling limit, where choosing the ghost-free structure raises the strong coupling scale from $\Lambda_5=\(\mpl m^4\)^{1/5}$ to $\Lambda_3=\(\mpl m^2\)^{1/3}$ \cite{ArkaniHamed:2002sp,Creminelli:2005qk}. However, just as the ghost-free structure of the potential changes the  leading operators from higher order scalar self-interactions \cite{Deffayet:2005ys} to ghost-free scalar-tensor Galileon interactions \cite{deRham:2010ik} in the flat case, in the decoupling limit on AdS the ghost-free structure ensures that the Proca interactions that arise at the scale $\Lambda_2$ are ghost-free. This will be seen more explicitly in what follows.

\paragraph{Ghost--free Proca on AdS:}

The Proca field $A_\mu$ has four components, yet only propagates the three degrees of freedom of a massive spin-1 at linear level.  This is due to the presence of the Proca constraint $\nabla_\mu A^\mu=0$ which is implied by the linear equations of motion.  At non-linear level, massive gravity with a generic potential has 6 degrees of freedom (the five of the massive graviton plus the Boulware Deser ghost \cite{Boulware:1973my}).
If we were to perform the massless AdS decoupling limit we are considering on a ghostly massive gravity potential, two of the degrees of freedom would go into the decoupled massless tensor, and so the remaining four must go into the interacting vector Lagrangian.  Thus we expect, in this generic case, that the Proca constraint would fail to hold at non-linear level.

However, the dRGT massive gravity theory \eqref{nonlinlag} is ghost free \cite{deRham:2010ik,Hassan:2011hr,Hassan:2011ea,deRham:2011rn,deRham:2011qq,Mirbabayi:2011aa,Hinterbichler:2012cn}, and so it possesses only five degrees of freedom.   Therefore in this case, the interacting Proca theory must carry only three degrees of freedom, and so must exhibit a Proca constraint at the fully non-linear level.  This is the way in which the massless decoupling limit of ghost-free dRGT on AdS will be different from that of a generic ghostly potential.

Interacting Proca theories which propagate three degrees of freedom non-linearly have been studied recently \cite{Tasinato:2014eka,Heisenberg:2014rta,Hull:2014bga,Tasinato:2014mia,Allys:2015sht,Hull:2015uwa,Charmchi:2015ggf,Jimenez:2016isa,Allys:2016jaq}.
However, we can see that the interactions \eqref{eq:Proca3p}, \eqref{eq:Proca4} we have obtained in this AdS decoupling limit of ghost-free massive gravity are not of the form of the generalized Proca interactions of \cite{Heisenberg:2014rta}.  Even after performing any local field redefinition on the vector field $A_\mu$, one can see that the interactions \eqref{eq:Proca4} cannot be put in a form that would match the generalized Proca theory, including its covariant version on AdS \cite{Heisenberg:2014rta}.

The reason the Proca interactions \eqref{eq:Proca3p}, \eqref{eq:Proca4} obtained through this decoupling limit of dRGT massive gravity do not belong to the class of interactions found in \cite{Heisenberg:2014rta} is that there it was asked that the constraint be satisfied at every order in perturbation theory in powers of the field\footnote{More precisely in powers of the second derivative of the helicity-0 mode of the massive vector field.}. In our decoupling limit on the other hand, there is an infinite number of interactions and truncating the theory at any finite order would spoil the existence of a constraint.

We can see how this works by computing the Hessian  of \eqref{fullProcaeseq} defined as  follows (see \cite{deRham:2011rn} for more details)
\be
{\cal H}={\rm det} \left| {\partial^2 {\cal L}\over \partial  \dot{ {A_\mu} }\partial   \dot{ {A_\nu} }}\right|\,,
\ee
where the dots indicates derivatives with respect to some time coordinate $t$ on AdS.
To be concrete, we can choose for instance Poincar\'e patch coordinates on $AdS_4$, $x^\mu=\{t,x,y,z\}$, in which the metric reads $\d s^2={L^2\over z^2}\left(-\d t^2+\d x^2+\d y^2+\d z^2\right)$.
If the Hessian does not vanish, then there are no constraints in the Hamiltonian formulation and the Lagrangian propagates four degrees of freedom.  Thus the Hessian must vanish to have the required constraint.  Using our explicit Lagrangian, we can compute the Hessian in a power series in $\Lambda_2^{-2}$.  Up to order $\Lambda_2^{-4}$, it does indeed vanish for any choices of parameters $\alpha_{3,4}$,
\be
{\cal H}=0+{\cal O}\left({1\over \Lambda_2^6 }\right),
\ee
and it does so through very non-trivial cancellations between the various orders in the Lagrangian, i.e. the Hessian of $\mathcal{L}^{(3)}_{\Lambda_2}$ and the Hessian of $\mathcal{L}^{(4)}_{\Lambda_2}$ do not individually vanish. Had the operators that entered the cubic or quartic Lagrangians been ever so slightly different, as they would be for a ghostly graviton potential, the cancellation of the Hessian would have been spoiled already at that order.

Since the full massive gravity is ghost free, we must have ${\cal H}=0$ to all orders, as well as an associated secondary constraint in the Hamiltonian formalism.  This is similar to the way in which the constraint appears which removes the Boulware-Deser ghost in dRGT theory in the unitary gauge \cite{Hassan:2011hr,Hassan:2011ea}.   It would be interesting to see precisely how the primary and secondary constraints manifest themselves at all orders in the AdS decoupling limit Proca theory.
By itself, the Proca theory can be seen as a special example of non-linear sigma model as proposed in \cite{deRham:2015ijs}.

\paragraph{AdS Proca/CFT duality:} At this point it is worth pointing out that we expect massive gravity on AdS to have a {large $N$ CFT dual with a non-conserved spin-2 single-trace primary}. It has also been suggested \cite{Vegh:2013sk,Blake:2013bqa,Blake:2013owa,Davison:2013jba} that massive gravity on AdS is dual to a condensed matter system or a CFT on a lattice, where the breaking of translation invariance on the boundary can be linked to the breaking of diffeomorphism invariance in the bulk.

With this picture in mind, a natural question is therefore whether the AdS decoupling limit of massive gravity is also dual to a particular limit of the CFT.  This would be a limit in which the dimension of the spin-2 primary approaches its conserved value $\Delta_c=3$, and the divergence of the current decouples to become its own spin-1 primary whose correlators decouple from the tensor correlators.  This would imply the existence of a new type of limit of the AdS/CFT correspondence where the boundary maps not to a gravitational theory but rather to a highly interacting Proca theory.  Furthermore, the theory which comes from the ghost-free potential should have special properties. These possibilities will be explored elsewhere.

\subsection{Flat spacetime limit}

Once we have the massless AdS decoupling limit, we can make contact with the flat space decoupling limit with $H/m\rightarrow 0$ by taking a further flat space limit $L\rightarrow \infty$.   In the flat limit, the mass of the vector field goes to zero and the degrees of freedom of the massive vector break up into a massless vector and a scalar, so we must now introduce the scalar \stu field and associated $U(1)$ symmetry to capture the dynamics of the scalar,
\be
A_\mu\rightarrow A_\mu+L\nabla_\mu \phi\,.
\ee
The scaling with $L$ will ensure that $\phi$ comes out canonically normalized.

We now have a new $U(1)$ symmetry
\bea \delta A_\mu=\partial_\mu\zeta ,\ \ \ \delta\phi=-{1\over L}\zeta\,, \label{massivenolimitstukmassless}
\eea
with gauge parameter $\zeta$.
At the linear level, after taking the limit $1/L \rightarrow 0$ while keeping both canonically normalized fields $A_\mu$ and $\phi$ fixed, the linearized Lagrangian \eqref{linearzedlagliste} becomes that of a massless vector and massless scalar on flat space,
\begin{equation}
{\cal L}=-\frac{1}{2} {F}_{\mu \nu}{F}^{\mu \nu}-6 \(\p \phi\)^2\,.
\end{equation}
The $U(1)$ symmetry \eqref{massivenolimitstukmassless} in the limit becomes
\bea \delta A_\mu=\partial_\mu\zeta ,\ \ \ \delta\phi=0, \label{massivenolimitstukmassless2}
\eea
and we see that $\phi$ is invariant.

Non-linearly, the flat decoupling limit is achieved by taking
 \be
 \Lambda_2\rightarrow \infty,\ \ \ {1\over L}\rightarrow 0,\ \ \ \Lambda_{3}=(\Lambda_2^2/L)^{1/3} \ \ {\rm fixed}.
 \label{lambda2flatdeclimite}\,
 \ee
A naive power counting would seem to suggest the existence of cubic interactions that enter at the strong coupling scale $\Lambda_5=(\Lambda_2^2L^{-3})^{1/5}$, however ghost-free massive gravity carries interactions in such a way that the cubic scalar operators that would enter at that scale $\(\p^2\phi\)^3/\Lambda_{5}^5$  always come in total derivative combinations. The same remains true to all orders in $\phi$ and the smallest suppression scale that remains is $\Lambda_{3}=(\Lambda_2^2/ L)^{1/3}$.

The full scalar-vector Lagrangian up to quartic order in the fields in this limit is 
\ba\nn
\L&=&-\frac{1}{2} F\mn^2- 6 (\p \phi)^2 \\ \nn
 &+& \frac 1{\Lambda_3^3}\Big((1+6\alpha_3)F_\mu{}^\alpha F_{\nu \alpha} \Pi^{\mu\nu} -
(1+3\alpha_3)F_{\mu\nu}F^{\mu\nu} [\Pi] \Big)+\frac{12}{\Lambda_3^3}(1+4\alpha_3)\L_{{\rm gal},3}^{H=0}(\phi) \\ \nn
 &+&
\frac{2}{\Lambda_3^6}F^{\mu\alpha}F^{\nu}{}_\alpha  \bigg[
-\frac{3}{2}(\alpha_3+4 \alpha_4)X^{(2)}\mn[\Pi]
-(1-12 \alpha_4)[\Pi]\Pi\mn+\frac 12(1-24 \alpha_4)\Pi^2\mn
\bigg]\\
&+&  \frac{(1-24\alpha_4)}{\Lambda_3^6}
F^{\alpha \mu}F^{\beta \nu}\Pi_{\alpha \beta}\Pi_{\mu \nu}
  -\frac{8}{\Lambda_3^6}(1-{{12}}\alpha_4)\L_{{\rm gal},4}^{H=0}(\phi) \,,\nn\\ \label{eqapp:LH0}
\ea
where $\Pi\mn=\p_\mu \p_\nu \phi$, brackets denote traces, the tensors $X^{(n)}\mn$ are defined in Appendix \ref{app:GFapp} and the Galileon terms $\L_{{\rm gal},n}^{H=0}$ are those of Appendix \ref{dsGalileonapps} in the flat space limit.
After suitable field re-definitions, the vector-scalar interactions here match those obtained in \cite{Ondo:2013wka} to fourth order in the fields, as well as those obtained at cubic order in \cite{deRham:2010gu} for a particular example of massive gravity \cite{Gabadadze:2009ja,deRham:2009rm}, and the Galileon interactions should match \cite{deRham:2012kf} in the massless limit after continuing to AdS.

The equations of motion from \eqref{eqapp:LH0} are not second order, but we know that there cannot be ghosts since it comes from ghost free massive gravity.  If the invertible field redefinition $\phi\to\phi+\frac{1}{24\Lambda_3^3}F\mn F^{\mu\nu}$ is performed, the resulting Lagrangian has second order equations to cubic order.  The problem is then pushed to quartic order and one should account for all the quartic interactions to draw any conclusion.
Providing the explicit field redefinition that would make the full decoupling limit of massive gravity either on flat space time or on AdS free of higher derivatives and manifestly ghost-free is beyond the scope of this work, but these decoupling limits provide yet more examples of effective field theories that exhibit higher derivatives and yet are free of any type of Ostrogradsky-like ghost instability.
This is a typical example of how higher order equations of motion may still be compatible with the absence of ghosts when multiple fields are involved, as was shown in \cite{deRham:2011rn,deRham:2011qq} and more recently in \cite{Motohashi:2016ftl,Motohashi:2018pxg} (the same type of phenomena is also observed within the context of Horndeski theories \cite{Gleyzes:2014qga,Langlois:2015skt,Deffayet:2015qwa,Achour:2016rkg,Crisostomi:2016tcp,Crisostomi:2016czh,Langlois:2015cwa,deRham:2016wji
}, and in the decoupling limit of various Galileon extensions of massive gravity \cite{Gabadadze:2012tr,Andrews:2013ora}).

\section{Non-Linear Partially Massless Decoupling Limits}
\label{sec:PM}

We now turn to massive gravity on de Sitter and consider the fully non-linear partially massless decoupling limit where, as depicted on Fig.~\ref{plot1}, we force ourselves to be arbitrarily close to the partially massless line.

No matter what the interactions present in the potential and kinetic term of the massive spin-2 field are (so long as they only involve at most two derivatives), there is no full non-linearly realized partially massless symmetry among ghost-free models \cite{deRham:2013wv,Garcia-Saenz:2014cwa}. The theory of massive gravity with the Einstein-Hilbert kinetic term, the cosmological constant and the dRGT potential interactions at $m^2=2H^2$  is the closest one can get to a partially massless candidate \cite{deRham:2012kf,deRham:2013wv}, but even with that theory in mind, there is no choice of parameters for which the symmetry would be restored \cite{deRham:2013wv}.

The absence of non-linear partially massless theory is not an issue in considering the partially massless limit, since the limiting theory keeps all  five degrees of freedom (just as the massless decoupling limit has all five degrees of freedom).   We know that as we approach the partially massless mass, $m^2\rightarrow 2H^2$, one of the modes will become strongly coupled (in the Vainshtein sense).  This is for the same reasons that the $m\rightarrow 0$ limit becomes strongly coupled; the \stu fields carrying the extra degrees of freedom and restoring the broken gauge symmetry becomes strongly interacting as we approach the enhanced gauge symmetry line. This strong coupling\footnote{By  `strong coupling' we do not imply here that the coupling constant $g_*\gg 1$, see \cite{deRham:2017xox} for more discussions on that point.} is precisely what ensures that the additional modes actually decouple in their respective massless or partially massless limit.   In the partially massless case, there is a \stu scalar carrying one extra degrees of freedom which restores partially massless symmetry away from $m=2H^2$, and which we will see becomes strongly coupled as $m^2\rightarrow 2H^2$.

For later convenience, it will be useful to point out that in the partially massless limit \eqref{pmdeclimlineare}, the linear theory \eqref{linearpmlimitdece} can be written in terms of two quantities that we now introduce.  The first is the partially massless field strength tensor \cite{Deser:2006zx},
\ba
\label{eq:FieldStrength}
F_{\mu\nu \lambda} \equiv \nabla_{\mu}h_{\nu \lambda}-\nabla_{\nu}h_{\mu \lambda}\, .
\ea
This quantity has mixed symmetry; it is anti-symmetric in the first two indices and vanishes if it is anti-symmetrized over all its indices.  It is manifestly invariant under the linear partially massless symmetry on de Sitter \eqref{partialmasslesssym}, $\delta_\chi F_{\mu\nu\lambda}=0$, and plays a role analogous to the Maxwell field strength in electromagnetism \cite{Hinterbichler:2014xga,Hinterbichler:2015nua}.
The second is the combination
\ba
\label{eq:phimn}
\Phi\mn \equiv \nabla_\mu \nabla_\nu  \phi+H^2 \phi\gamma\mn\,,
\ea
which is manifestly  invariant under extended Galileon-like shift symmetries for de Sitter space $\delta_B \Phi_{\mu\nu}=0$ (see appendix \ref{dsGalileonapps}).
In terms of these quantities, the linear action in the partially massless limit \eqref{linearpmlimitdece} is
\be \L^{\Delta \to 0}_2={\cal L}_{m_{\rm PM}}+{\cal L}_{\phi}\, , \ee
where
\ba
\label{eq:L2d}
{\cal L}_{m_{\rm PM}}(h)&=&\sqrt{-\gamma}\left(-\frac 14 F_{\mu\nu \lambda}F^{\mu\nu \lambda}+{1\over 2} F_{\mu\nu}^{\ \ \nu} F^{\mu\lambda}_{\ \ \ \lambda}\right) \, ,\\
{\cal L}_{\phi}&=&\sqrt{-\gamma}\frac{1}{2H^2} \Phi^{\mu\nu}X^{(1)}\mn(\Phi) \\
\nn
&=&\sqrt{-\gamma}\(-\frac 32 (\p \phi)^2 +6H^2 \phi^2\)=3{\cal L}_{{\rm gal},2}(\phi)\,.
\ea
Here $X^{(1)}\mn$ is
one of a family of tensors $X^{(n)}\mn$ defined in Appendix~\ref{app:GFapp}, and ${\cal L}_{{\rm gal},2}$ is one of the de Sitter Galileons, reviewed in Appendix \ref{dsGalileonapps}.

\subsection{Partially massless \stu}

At linear level, the \stu field $\phi$ is introduced as in \eqref{PMlinestue}, and restores the partially massless symmetry \eqref{PMstymsrese} away from $m^2=2H^2$.
Beyond linear level, we have a choice as to how to introduce the \stu field.  Since the theory at $m^2=2H^2$ has no partially massless symmetry beyond linear level, there is no natural choice for restoring it beyond linear level.  Here we will take inspiration from the vielbein formulation; in the vielbein language, the potential interactions for the massive graviton take a simple form \cite{Hinterbichler:2012cn} and truncate at a finite order in perturbation theory \cite{Gabadadze:2013ria}. Motivated by this, we will keep working in the metric language in what follows but with a vielbein-`inspired'
  choice of variables, where the metric is expressed as
      \ba
      \label{eq:defh}
      g\mn=\gamma^{\alpha\beta}(\gamma_{\mu \alpha}+{1\over\mpl}h_{\mu \alpha})(\gamma_{\nu \beta}+{1\over\mpl}h_{\nu \beta})\,.
      \ea
      This is equivalent to choosing $h_{\mu\nu}$ to be the fluctuation of the vielbein in a symmetric gauge.
  We then introduce the fully non-linear \stu field through the replacement
      \ba
      h\mn \to h\mn +\frac{1}{H \Delta }\Phi\mn\,,       \label{eq:defh2}
      \ea
where $\Phi\mn$ is as defined in \eqref{eq:phimn}.
This is the same form as the linear replacement \eqref{PMlinestue}.
The fully non-linear partially massless symmetry introduced in this way is thus identical to its linear version \eqref{PMstymsrese}\footnote{Note that $\phi$ is not invariant under the partially massless transformation before taking any decoupling limit, and neither should it be since it is a \stu field.  Only in the decoupling limit is $\phi$ invariant under the partially massless symmetry, and can be identified as a true scalar. The similarities that $\phi$ shares with scalars in this limit do not imply that $\phi$ is a scalar in the covariant sense of the term.}.

The \stu field $\phi$ as introduced above always enters in the combination $\Phi\mn$ defined in \eqref{eq:phimn}.  $\Phi\mn$ is not invariant under the standard global Galileon extended shift transformations, but rather it is invariant under the global de Sitter Galileon version of these transformations as introduced in \cite{Goon:2011uw,Burrage:2011bt,Goon:2011qf} (see Appendix \ref{dsGalileonapps}).  These transformations reduce to the standard Galileon and shift symmetries in the flat spacetime limit $H\to 0$.

At the linear level, we saw in Section \ref{PMlimitsec} that $h\mn$ decouples from $\phi$ in the partially massless limit $\Delta \to 0$.  However the absence of a non-linear partially massless theory indicates that at the non-linear level there will be higher order operators involving $\phi$ that do not vanish in the limit $\Delta \to 0$.  We therefore expect interactions that scale like inverse powers of $\Delta$ ($\Delta$ plays the same role here as the mass $m$ does in the massless decoupling limit.)

With our choice of  \stu field  \eqref{eq:defh2}, the Einstein--Hilbert term brings in an infinite number of interactions involving the scalar whereas with the structure \eqref{eq:defh} the dRGT mass terms only bring in a finite number (this is in contrast to what occurs when restoring diffeomorphism invariance in the massless limit, in which the \stu fields do not contribute through the Einstein--Hilbert term).  We will see that the resulting  decoupling limits only carry a finite number of pure-$\phi$ interactions, which we will find to be the de Sitter Galileon interactions derived in \cite{Goon:2011uw,Burrage:2011bt,Goon:2011qf}.

The interactions we get from the \stu replacement \eqref{eq:defh}, \eqref{eq:defh2} are of the following schematic form:
\ba
\label{eq:Lelln}
\L_{\ell, n} &\sim & (\p+H)^{2n+2} \frac{1}{\mpl^{\ell-2}}\frac{1}{(H\Delta \mpl)^n} h^\ell \phi^n\,, \ \ n,\ell\geq 0,\ \ n+\ell\geq 3\,. \\
\tilde \L_{\ell, n} &\sim & (\p+H)^{2n} \Delta^2 \frac{1}{\mpl^{\ell-2}}\frac{1}{(H\Delta \mpl)^n} h^\ell \phi^n\,, \ \ n,\ell\geq 0,\ \ n+\ell\geq 3\,. \label{eq:Lelln2}
\ea
(Here we treat $\partial$, $H$ as schematically identical, since they are held fixed in the limit of interest and can be converted into each by commuting derivatives.)  The terms \eqref{eq:Lelln} come from the Einstein-Hilbert and mass terms upon the \stu substitution.  The terms \eqref{eq:Lelln2}, with the extra $\Delta^2$ suppression, come from the mass term when we replace $m^2\rightarrow \Delta^2+2H^2$; they are the terms proportional to $\Delta^2$.  This $\Delta^2$ suppression will play an important role later.

The strong coupling scale will be the smallest scale appearing among the terms \eqref{eq:Lelln}, \eqref{eq:Lelln2}, and this is the scale that will be held fixed in the decoupling limit.  We define the following analogues of the massive gravity scales,
\be \tilde \Lambda_{k}=\left(\mpl \Delta^{k-1}\right)^{1/k}\, ,\ee
and we sort the various terms \eqref{eq:Lelln}, \eqref{eq:Lelln2} according to which $\tilde \Lambda_{k}$ scale they appear with.

 We recall that $\phi$ is introduced as a \stu field to restore the partially massless symmetry in the full theory. In the decoupling limit the \stu field and the graviton decouple (at least partially) and the \stu field then transforms in that limit as a scalar under the partially massless transformation. This implies that in the decoupling limit, the resulting theory should be  manifestly invariant under the linear partially massless transformation \eqref{PMstymsresel}.  We also expect it to be invariant under the de Sitter version of Galileon symmetry reviewed in Appendix \ref{dsGalileonapps}, since the \stu is introduced through the dS Galileon invariant $\Phi\mn$ defined in \eqref{eq:phimn} (just as the flat space decoupling limit is invariant under ordinary Galileon symmetry).  We thus expect the resulting action to be built out of the partially massless invariant field strength $F_{\mu\nu\lambda}$ defined in \eqref{eq:FieldStrength}, and the dS Galileon invariant $\Phi\mn$ defined in \eqref{eq:phimn}.

\subsection{$\tilde \Lambda_4$ decoupling limit\label{lambda4subss}}

Generically, i.e. if the parameters $\alpha_3,\alpha_4$ are chosen arbitrarily, the lowest scale appearing among the interactions \eqref{eq:Lelln}, \eqref{eq:Lelln2} is $\tilde\Lambda_4=\(\mpl \Delta^3\)^{1/4}$, carried by the terms $\L_{0, 3}$ which are cubic $\phi$ self-interactions.
 The terms carrying this scale that survive have up to four derivatives (all the interactions with more derivatives appear as a total derivative combinations, due to the ghost-free structure of the dRGT Lagrangian.  The same will be true for terms higher order in the fields.) The decoupling limit is taken by sending  $\Delta \to 0$, $\mpl \to \infty$ while keeping the scale $\tilde\Lambda_4=\mpl \Delta^3$ (and the dS scale $H$ as well as all the canonically normalized fields) fixed,
\be
\mpl \rightarrow \infty,\ \ \ \Delta\rightarrow 0,\ \ \ \tilde\Lambda_4, \ H \ \ {\rm fixed},\ \ \ h_{\mu\nu},\ \phi \ {\rm fixed}\,.
\label{lambda42pmdeclimite}
\ee
The resulting Lagrangian in this limit is
\be {\cal L}_{\tilde \Lambda_4}={\cal L}_{m_{\rm PM}}(h)+ 3 {\cal L}_{{\rm gal},2}(\phi)+{3H(1+2\alpha_3)\over \tilde\Lambda_4^4} {\cal L}_{{\rm gal},3}(\phi)\, .\label{lambda32declage}
\ee
Here, in addition to the kinetic terms of the linearized theory, we see that the cubic interaction terms have organized themselves into ${\cal L}_{{\rm gal},3}$, which is the cubic de Sitter Galileon discovered in the context of brane world constructions \cite{deRham:2010eu} in \cite{Goon:2011uw,Burrage:2011bt,Goon:2011qf} (see Appendix \ref{dsGalileonapps} for definitions).
The strong coupling scale is
\be
\Lambda= \left(\tilde\Lambda_4^4/H\right)^{1/3}=\left(\mpl \Delta^3/H\right)^{1/3} \,.
\ee
As expected, the decoupling limit Lagrangian is manifestly invariant under the linear partially massless symmetry \eqref{PMstymsresel}, as well as the extended de Sitter Galileon shift symmetries reviewed in Appendix \ref{dsGalileonapps}.

\subsection{$\tilde\Lambda_{3}$ decoupling limit\label{lambda3subss}}

Looking at \eqref{lambda32declage}, we see that in the case where $\alpha_3=-1/2$, the cubic Galileon interactions in the decoupling limit \eqref{lambda42pmdeclimite} vanish.  This means that for this choice of $\alpha_3$ the true strong coupling scale is higher.  The next highest scale appearing among the interactions \eqref{eq:Lelln}, \eqref{eq:Lelln2} is $\tilde\Lambda_3=\(\mpl \Delta^2\)^{1/3}$, carried by the terms $\L_{0, 4}$ which are quartic $\phi$ self-interactions, as well as the terms $\L_{1, 3}$ which are linear in $h$.

This $\tilde\Lambda_3$ decoupling limit is then
\be
\mpl \rightarrow \infty,\ \ \ \Delta\rightarrow 0,\ \ \ \tilde\Lambda_3,\ H \ \ {\rm fixed}\, ,\ \  \ h_{\mu\nu},\ \phi\ \ {\rm fixed}\,,
\label{lambda3pmdeclimite}
\ee
keeping again the canonically normalized fields and the scale $H$ fixed in the limit.

As we will explain is Section \ref{fulldeclims}, the terms in $\L_{1, 3}$ (that are not suppressed by additional powers of $\Delta^2$) vanish identically up to total derivatives.  This leaves only the $\L_{0, 4}$ quartic $\phi$ self interactions, and the
Lagrangian in the limit \eqref{lambda3pmdeclimite} is
\ba
{\cal L}_{\tilde \Lambda_3}={\cal L}_{m_{\rm PM}}( h)+ 3{\cal L}_{{\rm gal},2}( \phi)+{8\alpha_4-1 \over 2\tilde\Lambda_3^6} {\cal L}_{{\rm gal},4}( \phi)\, .\label{lambda42declage}
\ea
The $\phi^4$ terms take the form of a quartic de Sitter Galileon, as defined in Appendix \ref{dsGalileonapps}.  The strong coupling scale $\Lambda$ is simply $\tilde \Lambda_3$,
\ba
\Lambda= \tilde \Lambda_3=\left(\mpl \Delta^2\right)^{1/3}\,.
\ea

\subsection{$\tilde\Lambda_{2}$ decoupling limit\label{fulldeclims}}

Looking at \eqref{lambda42declage}, we see that in the case where
\ba
\label{eq:alphas}
\alpha_3=-1/2\, , \qquad \alpha_4=1/8\,,
\ea
the quartic Galileon interactions in the decoupling limit \eqref{lambda3pmdeclimite} vanish.  This means that in this case the true strong coupling scale is higher yet.   In fact, the values \eqref{eq:alphas} are precisely the values of the ``candidate partially massless theory" identified in \cite{deRham:2012kf} and studied in \cite{deRham:2013wv}.  These are the values for which, in some sense, the dRGT theory comes closest to realizing a full partially massless symmetry.  The theory at this point does not have full partially massless symmetry, but it has partially massless symmetry to cubic order in the interactions, to all orders in the scalar-tensor sector of the flat space decoupling limit, and in FRW cosmological solutions and fluctuations about them.  In addition, it has a $Z_2$ symmetry between the dynamical metric and the background metric.
We will refer to the values \eqref{eq:alphas} as the partially massless values.

The fact that at the partially massless values \eqref{eq:alphas} the scalar-tensor interactions vanish in the flat space decoupling limit means that all the way up to the scale $\Lambda_3=(\mpl m^2)^{1/3}$ in this limit, the partially massless symmetry is preserved \cite{deRham:2012kf}. This means that upon choosing the appropriate partially massless values \eqref{eq:alphas} for the parameters $\alpha_3$ and $\alpha_4$,
all the pure $\phi$ interactions as well as all the interactions that are linear in $h\mn$ should be suppressed with at least one power of $\Delta^2$, \ie with the appropriate choice of parameters $\alpha_{3,4}$, the interactions $\L_{0, n}$ and $\L_{1, n}$ from \eqref{eq:Lelln} should vanish.
This suppression has profound consequences for the value of the strong coupling scale, since these are the only terms with scales smaller than $\LPM=\left(\mpl \Delta\right)^{1/2}$.   Once we see that this suppression occurs, i.e. the terms $\L_{0, n}$, $\L_{1, n}$ defined in  \eqref{eq:Lelln} vanish, the next highest scale is $\LPM$, carried by the terms $\L_{2, n}$, which are quadratic in $h$, and the terms $\tilde \L_{0, n}$ defined in \eqref{eq:Lelln2} which are scalar self-interactions suppressed by $\Delta^2$.

Following the previous arguments we are lead to a new and interesting decoupling limit of massive gravity, which can only be taken when the parameters take the partially massless values \eqref{eq:alphas}.  In this limit we take $\mpl \to \infty$, keep the Hubble constant fixed and send $\Delta^2=m^2-2H^2\to 0$ keeping the scale $\LPM=(\mpl \Delta)^{1/2}$ fixed,
\be
\mpl \rightarrow \infty,\ \ \ \Delta\rightarrow 0,\ \ \ \tilde\Lambda_2,\ H \, \ {\rm fixed} ,\ \ \  h_{\mu\nu},\ \phi\  \ {\rm fixed}\,.
\label{lambda22pmdeclimitee}
\ee
With these considerations in mind we now derive the full form of the $\LPM$ decoupling limit.  As already mentioned, this decoupling limit includes terms which are at most quadratic order in $h\mn$.  We hence proceed order by order in $h\mn$, first showing the vanishing of the terms $\L_{1, n}$ linear in $h\mn$ and the suppression by $\Delta^2$ of the scalar self-interactions (i.e. showing vanishing of $\L_{0, n}$ and deriving $\tilde \L_{0, n}$), and then deriving the terms $\L_{2, n}$ quadratic in $h$.

\paragraph{Up to linear order in $h\mn$:}  The beauty of the vielbein inspired choice \eqref{eq:defh}, \eqref{eq:defh2} for the introduction of the \stu scalar is that the Einstein-Hilbert and cosmological constant terms (which are simple wedge products in the vielbein formulation)
contribute up to first order in $h\mn$ only a finite number of terms.  These terms are
\ba
\L_{\rm EH}=\frac{\mpl^2}{2}\sqrt{-g}\(R[g]-\Lambda\)&= &
 - \frac{2H}{\Delta}  \Bigg[
\(h^{\mu\nu}+\frac 1{2H \Delta}  \Phi^{\mu\nu}\)X^{(1)}\mn(\Phi)\\
&+&\frac{3}{4H \LPM^2}\( h^{\mu\nu}+\frac 1{3H\Delta}  \Phi^{\mu\nu}\)X^{(2)}\mn(\Phi)\nn\\
&+&\frac{1}{4H^3 \LPM^4}\(h^{\mu\nu}+\frac 1{4H\Delta}  \Phi^{\mu\nu}\)X^{(3)}\mn(\Phi)
\Bigg]+\mathcal{O}(h^2)\,.\nn
\ea
Here the $X^{(n)}\mn$ are given in Appendix \ref{app:GFapp}.  This expression is exact to all orders in $\phi$.

As for the contributions from the mass terms in \eqref{nonlinlag}, we know that they are simple polynomials in terms of vielbeins \cite{Hinterbichler:2012cn}, and so there are also a finite number of contributions to first order in $h$,
\ba
\L_{\rm mass}&=&{\mpl^2m^2 \over 2}\sqrt{-g}\left[S_2({\cal K})-{1\over 2} S_3({\cal K})+{1\over 8} S_4({\cal K})\right]=
  \frac{m^2}{H\Delta}  \Bigg[
\(h^{\mu\nu}+\frac 1{2H \Delta}  \Phi^{\mu\nu}\)X^{(1)}\mn(\Phi) \nn\\
&+&\frac{3}{4H \LPM^2}\( h^{\mu\nu}+\frac 1{3H\Delta}  \Phi^{\mu\nu}\)X^{(2)}\mn(\Phi)
+\frac{1}{4H^3 \LPM^4}\(h^{\mu\nu}+\frac 1{4H\Delta}  \Phi^{\mu\nu}\)X^{(3)}\mn(\Phi)
\Bigg]\nn\\
&+&\mathcal{O}(h^2)\,.
\ea

Individually, the Einstein-Hilbert and mass terms would seem to blow up in the $\LPM$ decoupling limit where $\Delta\to 0$ keeping $\LPM$ fixed.  However, with the choice of parameters \eqref{eq:alphas}, and the mass $m^2=2H^2+\Delta^2$, we see that the terms coming from the $2H^2$ part of the $m^2$ will precisely cancel against the Einstein--Hilbert contribution.  This leaves everything else suppressed by $\Delta^2$.  The terms linear in $h$ with this extra $\Delta^2$ suppression contribute at a higher scale than $\LPM$ so we can ignore them.   We are thus left with the pure $\phi$ terms at the scale $\LPM$, and these take the form of de Sitter Galileons,
\ba
\L_{\rm PM} &=&  \L_{\rm EH} + \L_{\rm mass}
\to \L_{\rm gal} +\L_{ h^2}\,,
\ea
where the de Sitter Galileon interactions are given by
\ba
\L_{\rm gal} &=&
\frac{1}{2H^2}  \Phi^{\mu\nu} X^{(1)}\mn(\Phi)
+{1\over 4H^3\LPM^2}  \Phi^{\mu\nu} X^{(2)}\mn(\Phi)
+{1\over 16H^4\LPM^4}  \Phi^{\mu\nu} X^{(3)}\mn(\Phi) \nn
\\
&=&3\, {\cal L}_{{\rm gal},2}( \phi)+{3\over 2H\LPM^2} {\cal L}_{{\rm gal},3}( \phi)+{1\over 4H^2\LPM^4} {\cal L}_{{\rm gal},4}( \phi)\,\,.  \label{lambda2Galileonse}
\ea
They have the de Sitter version of Galileon symmetry, as reviewed in Appendix~\ref{dsGalileonapps}.

\paragraph{Quadratic order in $h\mn$:}
The final part of the $\LPM$ partially massless decoupling limit is given by the terms quadratic in $h\mn$.  These $h^2$ terms must have the partially massless symmetry
\eqref{PMstymsresel}, and so we expect them to be writable in terms of the partially massless invariant field strength \eqref{eq:FieldStrength}.

Computing the explicit contributions from the $h^2$ terms is relatively straight-forward but requires expanding the Einstein--Hilbert term and the mass term to second order in $h\mn$ about a generic metric $V\mn$ which is then identified to $\gamma\mn+ \frac{1}{H \LPM^2}\Phi\mn$ (in our vielbein-inspired variables). This is straightforward exercise, and after some integrations by parts we are left with
\bea
\L_{h^2}&=&-{1\over 4}\sqrt{-g}\left|\det V\right|\bigg[{1\over 2}F_{\mu\alpha\rho}F_{\nu\beta\sigma}\left(V^{-2}\right)^{\mu\nu}\left(V^{-2}\right)^{\alpha\beta}\gamma^{\rho\sigma}  \nn\\
&& - \left(2F_{\mu\rho\sigma}F_{\nu\alpha\beta}-F_{\mu\alpha\rho}F_{\nu\sigma\beta}\right)\left(V^{-2}\right)^{\mu\nu}\left(V^{-1}\right)^{\alpha\beta}\left(V^{-1}\right)^{\rho\sigma}\bigg] \,,\label{h2phidecoe}
\eea
with $F_{\mu\nu\lambda}$ the partially massless field strength tensor defined in \eqref{eq:FieldStrength} and
\ba
V\mn&\equiv&\gamma\mn + \frac{1}{H \LPM^2}\Phi\mn= \gamma\mn+\frac{1}{H\LPM^2}\left( \nabla_\mu \nabla_\nu \phi + {H^2} \phi \gamma\mn\right)\,, \label{Vdefe}
\ea
with $\left(V^{-1}\right)^{\mu\nu}$ the matrix inverse of $V_{\mu\nu}$, so that $\left(V^{-1}\right)^{\mu\lambda}V_{\lambda\nu}=\delta^{\mu}_{\nu}$.  The quantity $\left(V^{-2}\right)^{\mu\nu}\equiv \left(V^{-1}\right)^{\mu\rho}\left(V^{-1}\right)^{\nu\sigma}\gamma_{\rho\sigma}$ is nothing other than the full inverse metric evaluated at $h=0$, and $|\det V|=\sqrt{-g}|_{h=0}$.\\

The full decoupling limit is given by the sum of the Galileon terms in \eqref{lambda2Galileonse} and the $h^2$ terms in \eqref{h2phidecoe},
\bea
{\cal L}_{\tilde \Lambda_2}&=& 3\, {\cal L}_{{\rm gal},2}( \phi)+{3\over 2H\LPM^2} {\cal L}_{{\rm gal},3}( \phi)+{1\over 4H^2\LPM^4} {\cal L}_{{\rm gal},4}( \phi)\, \nn\\
&&-{1\over 4}\sqrt{-g}\left|\det V\right|\bigg[{1\over 2}F_{\mu\alpha\rho}F_{\nu\beta\sigma}\left(V^{-2}\right)^{\mu\nu}\left(V^{-2}\right)^{\alpha\beta}\gamma^{\rho\sigma}  \nn\\
&& - \left(2F_{\mu\rho\sigma}F_{\nu\alpha\beta}-F_{\mu\alpha\rho}F_{\nu\sigma\beta}\right)\left(V^{-2}\right)^{\mu\nu}\left(V^{-1}\right)^{\alpha\beta}\left(V^{-1}\right)^{\rho\sigma}\bigg] \,,\label{fullpmlagalle}
\eea
with $V\mn$ as defined in \eqref{Vdefe} and $F_{\mu\nu\lambda}$ as defined in \eqref{eq:FieldStrength}.  The strong coupling scale is
\ba
\Lambda= \( H\tilde \Lambda_2^2 \)^{1/3}=\left(H \mpl \Delta\right)^{1/3}\,.
\ea

The partially massless Lagrangian \eqref{fullpmlagalle} is manifestly invariant under the local partially massless symmetry \eqref{PMstymsresel}, since the field strength $F_{\mu\nu\lambda}$ is invariant under it. The Lagrangian is also invariant under the global de Sitter Galileon symmetry since $\Phi_{\mu\nu}$, and hence $V_{\mu\nu}$, is invariant under it.

\subsection{Scalar potential and ${\mathbb Z}_2$ symmetry}

It is clear that the decoupling limit carries a mild instability for the field $\phi$ with a time scale of order the Hubble parameter, as is already clear from the negative mass square term for $\phi$ in \eqref{linearpmlimitdece} at the linear level.  This instability would be problematic if we intended to treat the theory as a standard fundamental quantum field theory, however given that it is already an effective field theory, on a classical and phenomenological level the instability is not of great concern since the scale of the instability is of order the Hubble parameter.

\begin{figure}[h!]
\begin{center}
\epsfig{file=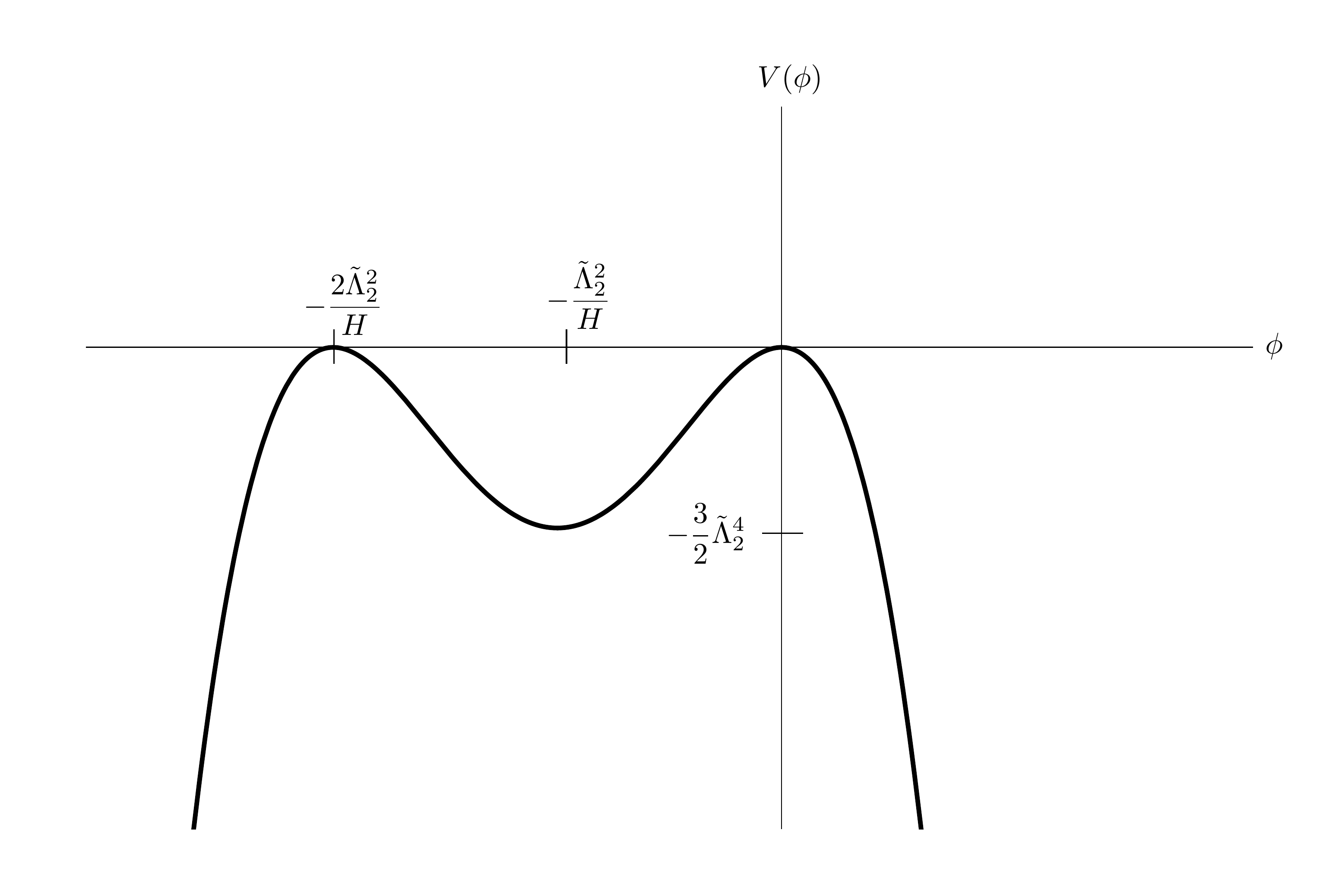,height=4.0in,width=5.8in}
\caption{Scalar potential in the $\tilde\Lambda_2$ decoupling limit.}
\label{plot2}
\end{center}
\end{figure}

Furthermore, this instability is simply a reflection of the Galileon shift symmetry, one of whose generators is realized in the flat-slicing inflationary patch of de Sitter, $\d s^2 = - \d t^2 +e^{2 H t} \d \vec{x}{}^2$, as an exponential growth $\delta \phi\propto e^{H t}$. Since this is a global symmetry of the theory, any point in the potential would carry the same feature. For instance the effective potential carries a local minimum as will be shown below (see Fig.~\ref{plot2}), and at that point the instability manifests itself instead through the kinetic term.  Since this `instability' is a consequence of the Galileon shift symmetry on de Sitter, it is only observable if the Galileon couples to external sources in a way that breaks the de Sitter Galileon symmetry. If instead, we consider configurations related by a Galileon shift symmetry as equivalent then the field $\phi$ itself would not be a gauge-invariant quantity and a more appropriate question is related to the behavior of correlation functions of a `gauge invariant' quantity like the `field strength` $\Phi\mn$ as defined in \eqref{eq:phimn}, which is insensitive to the mild instability seen by the field $\phi$.

To make manifest the ${\mathbb Z}_2$ symmetry of the  $\tilde \Lambda_2$ decoupling limit, we first point out that since the Lagrangian is quadratic in $h_{\mu\nu}$, in the vacuum there is a classically consistent truncation where we can set $h_{\mu\nu}=0$ in the action.  This leaves only the scalar self-interactions governed by the cubic and quartic dS Galileons.  These Galileon interactions include a cubic and quartic potential.  The full scalar potential, including the mass term from the quadratic Galileon reads
\be V(\phi)=-6 H^2 \phi ^2-\frac{6 H^3 \phi ^3}{\tilde \Lambda_2 ^2}-\frac{3 H^4 \phi ^4}{2 \tilde \Lambda_2 ^4}\,.\ee
This potential is a ${\mathbb Z}_2$ symmetric upside-down spontaneous symmetry breaking potential, expanded around one of its ${\mathbb Z}_2$ breaking maxima (see Fig.~\ref{plot2}).  The extrema are at
\bea && {\rm maxima:}\ \ \ \ \ \ \ \ \ \ \ \ \phi=0,\ -{2\tilde \Lambda_2 ^2\over H}  \nn\\
&& {\rm local\ minimum:}\ \ \ \phi=-{\tilde \Lambda_2 ^2\over H}\,.
\eea
We can make the ${\mathbb Z}_2$ symmetry manifest by expanding around the symmetry preserving local minimum,
\be \phi=\phi'-{\tilde \Lambda_2 ^2\over H} .\ee
The Lagrangian in terms of $\phi'$ is also of the Galileon form, and involves only the ${\mathbb Z}_2$ symmetric quadratic and quartic terms,
\be {\cal L}= -{3\over 2}{\cal L}_{{\rm gal},2}(\phi')+ {1\over 4H^2 \tilde\Lambda_2 ^4} {\cal L}_{{\rm gal},4}(\phi').\ee
Note that the kinetic term here has the wrong sign, so this local minimum is actually unstable due to a ghost instability in the kinetic terms.  Thus there is no truly stable configuration for the field $\phi$, which is a consequence of the de Sitter Galileon symmetry.

\section{Conclusions}
\label{sec:conc}

We have considered two kinds of decoupling limits of ghost-free massive gravity on curved space.   The first, a massless limit on fixed AdS, comes with a strong coupling scale $\Lambda_2=\left(\mpl m\right)^{1/2}$, and is described in the decoupling limit by a self-interacting massive Proca theory.  While the existence of this limit and the value of the strong coupling scale in this limit is well known, here we derived the explicit form of the interacting massive Proca theory up to quartic order in the fields, showing that it is not of the form of the ghost-free generalized Proca theories previously studied, yet still non-trivially maintains the constraint necessary to describe three degrees of freedom non-linearly.  The second limit we considered is a new type of partially massless decoupling limit on dS.  In this limit, the strong coupling scale can be raised, by choosing the partially massless parameters, to $\left(H \mpl \Delta\right)^{1/3}$, where $\Delta^2=m^2-2H^2$ measures the departure from the partially massless point.  The \stu field in this case is a scalar that restores partially massless symmetry.  We derived the exact all-orders decoupling limit in this case, finding that the scalar self interactions are governed by de Sitter Galileons.

The AdS limit is of interest because there are examples in which a light graviton mass is induced on AdS.  For example, in the construction of \cite{Porrati:2001db,Porrati:2003sa}, a small graviton mass can be induced on AdS through a loop effect when Einstein gravity is coupled to a scalar field with non-standard boundary conditions (brane-world, AdS/CFT and string theory realizations of similar systems and bi-gravity extensions have been extensively studied \cite{Karch:2000ct,Karch:2001cw,Porrati:2001gx,Aharony:2003qf,Duff:2004wh,Kiritsis:2006hy,Aharony:2006hz,Kiritsis:2008at,Apolo:2012gg,Bachas:2017rch,Bachas:2018zmb}).  In this case the graviton mass is $m\sim \(L^2 \mpl\)^{-1}$, and so the strong coupling scale is the same as the AdS scale $\Lambda_2\sim \sqrt{\mpl m}\sim  1/ L$.   In the decoupling limit we consider here, we would have a massive vector with a mass $m_A\sim 1/L$ which is of the same order as the strong coupling scale, so the decoupling limit effective theory may have limited usefulness in this case.
Constructions in which five dimensional AdS gravity is used to induce a non-local theory in four dimensions with a higher cutoff have also been of interest recently \cite{deRham:2006pe,Gabadadze:2014rwa,Gabadadze:2015goa,Domokos:2015xka,Gabadadze:2017jom}.

The partially massless dS limit is of interest because the formalism is now in place to understand the classical and quantum behavior of massive gravity close to the partially massless line, and to address questions such as how the Vainshtein mechanism is realized in the partially massless limit.  Within a late-time cosmological setup, it is not inconceivable that $\Delta$ is arbitrarily close to its partially massless value today and understanding the partially massless limit phenomenology could go a long way towards determining the viability of the model.  Alternatively, massive spin-2 fields could be present during inflation and contribute to primordial gravitational waves or other signatures \cite{Arkani-Hamed:2015bza,Kehagias:2017cym,Baumann:2017jvh,Biagetti:2017viz,Franciolini:2017ktv,Anber:2018bau,MoradinezhadDizgah:2018ssw,Franciolini:2018eno,MoradinezhadDizgah:2018pfo,Bordin:2018pca,Dimastrogiovanni:2018uqy}.  Stability requires these massive spin-2 fields to have a mass above the Higuchi bound, which is the partially massless line.   To have a large phenomenological impact, their mass should not be much higher than this bound at the time of inflation, and should thus be close to the partially massless point, exactly the regime captured by the partially massless decoupling limit.  We thus expect this limit to be useful in simplifying the study of such inflationary scenarios.

\bigskip
{\bf Acknowledgements:}
The authors are grateful to James Bonifacio, S\'ebastien Renaux-Petel and Andrew Tolley for discussions and comments.  CdR would like to thank Case Western Reserve University for its hospitality during PASCOS2018.
The work of CdR is supported by an STFC grant ST/P000762/1. CdR thanks the Royal Society for support at ICL through a Wolfson Research Merit Award. CdR is also supported in part by the European Union's Horizon 2020 Research Council grant 724659 MassiveCosmo ERC-2016-COG and in part by a Simons Foundation award ID 555326 under the Simons Foundation's Origins of the Universe initiative, `{\it Cosmology Beyond Einstein's Theory}'.  KH gratefully acknowledges support from the 2018 Simons Summer Workshop at the Simons Center for Geometry and Physics, Stony Brook University, at which some of the research for this paper was performed.

\appendix

\section{Symmetric polynomials and tensors}
\label{app:GFapp}

Here we define the symmetric polynomials $S_n$ and tensors $X^{(n)}\mn$ that enter throughout the paper.  They are the same objects that play a role in the standard $\Lambda_3$ decoupling limit of massive gravity \cite{deRham:2010tw,deRham:2014zqa}.

For an arbitrary symmetric tensor $M\mn$, we define the symmetric polynomials as
\be  S_n^{\rm }(M)=n!\, M^{[\mu_1}_{\ \mu_1}M^{\mu_2}_{\ \mu_2}\cdots M^{\mu_n]}_{\ \mu_n} \, .\ee
Explicitly, they are
\bea
\label{eq:S0}
 S_0({M}) &=&1, \\
\label{eq:S1}
 S_1({M}) &=&[{M}], \\
\label{eq:S2}
 S_2({M}) &=&[{M}]^2-[{M} ^2], \\
\label{eq:S3}
S_3({M})&=& [{M}]^3-3 [{M}][{M} ^2]+2[{M} ^3] ,\\
\label{eq:S4}
S_4 ({M})&=& [{M}]^4
-6[{M} ^2][{M}]^2+8[{M} ^3][{M}]+3[{M} ^2]^2 -6[{M} ^4] ,
\eea
where the brackets are matrix traces.

The tensors $X^{(n)}_{\mu\nu}(M)$ are defined as follows\footnote{Note that our definition of the $X^{(n)}_{\mu\nu}$ differs by a factor of 2 from that of \cite{deRham:2010kj}.},
\be X^{(n)\mu}_{\ \ \ \  \nu}(M)={1\over n+1}{\delta \over \delta M^\mu_{\ \nu}} S_{n+1}^{\rm }(M)=(n+1)!\, \delta^{[\mu}_{\ \nu}M^{\mu_2}_{\ \mu_2}\cdots M^{\mu_n]}_{\ \mu_n}.\ee
Explicitly, they are
\bea X^{(0)}_{\mu\nu}(M)&=&\gamma_{\mu\nu} \, ,\\
 X^{(1)}_{\mu\nu}(M)&=&\left[M\right]\gamma_{\mu\nu}-M_{\mu\nu} \, ,\\
  X^{(2)}_{\mu\nu}(M)&=&\left(\left[M\right]^2-\left[M^2\right]\right)\gamma_{\mu\nu}-2\left[M\right]M_{\mu\nu}+2M^2_{\mu\nu}\, , \\
   X^{(3)}_{\mu\nu}(M)&=&\left(\left[M\right]^3-3\left[M\right]\left[M^2\right]+2\left[M^3\right]\right)\gamma_{\mu\nu}-3\left(\left[M\right]^2-\left[M^2\right]\right)M_{\mu\nu}+6\left[M\right]M^2_{\mu\nu}-6M^3_{\mu\nu}\, .\nn \\
   \eea

The symmetric polynomials are proportional to the traces of these tensors,
\be X^{(n)\mu}_{\ \ \ \ \ \mu}(M)=(4-n)S_n(M)\,.
\ee

\section{de Sitter Galileons\label{dsGalileonapps}}

Here we review the de Sitter Galileons derived in \cite{Goon:2011uw,Burrage:2011bt,Goon:2011qf}, which make an appearance in the partially massless decoupling limit.  The de Sitter Galileons are similar to the flat space Galileons \cite{Nicolis:2008in} in that they are higher-derivative Lagrangians whose equations of motion are second order and can be derived using a probe brane embedding as proposed in \cite{deRham:2010eu}.   Unlike the covariantized Galileons \cite{Deffayet:2009wt}, they maintain an extended Galileon shift symmetry, and they have mass and potential terms whose form is fixed by the extended symmetry.

The precise normalization we use and explicit expressions for the de Sitter Galileons are
\begin{eqnarray}
 {\cal L}_{{\rm gal},1}(\phi)&=&\sqrt{-\gamma} \phi \ , \nn\\
 {\cal L}_{{\rm gal},2}(\phi)&=&\sqrt{-\gamma} \left[-\half(\partial \phi)^2+{2H^2}  \phi^2\right] \ ,\nn \\
 {\cal L}_{{\rm gal},3}(\phi)&=& \sqrt{-\gamma}\left[-{1\over 2}(\partial \phi)^2[ \Pi]-{3 H^2} (\partial \phi)^2 \phi+{4 H^4} \phi^3\right] \ ,\nn\\
 {\cal L}_{{\rm gal},4}(\phi)&=&\sqrt{-\gamma}\left[-\half(\partial \phi)^2\left([ \Pi]^2-[ \Pi^2]+{H^2\over 2}(\partial \phi)^2+{6H^2} \phi[ \Pi]+{18H^4} \phi^2\right)+{6 H^6} \phi^4\right] \ , \nn \\
 {\cal L}_{{\rm gal},5}(\phi)&=& \sqrt{-\gamma}\left[-\half\left((\partial \phi)^2+{H^2\over 5}  \phi^2\right)\left([ \Pi]^3-3[ \Pi][ \Pi^2]+2[ \Pi^3]\right)\right. \nn \\
&&\left.-{12H^2\over 5}  \phi(\partial  \phi)^2\left([ \Pi]^2-[ \Pi^2]+{27H^2 \over 12}[ \Pi]  \phi+{5 H^4}  \phi^2\right)+{24H^8\over 5}  \phi^5\right]  \ ,
\label{dsGalileonsscaled}
\end{eqnarray}
where $\Pi_{\mu\nu}=\nabla_\mu\nabla_\nu\phi$, and the brackets are traces.  They reduce to the flat space Galileons when $H\rightarrow 0$.

The de Sitter Galileons for $n\leq 4$  can be written up to a total derivative in terms of the tensors $X^{(n)}\mn$ defined in Appendix \ref{app:GFapp},
\bea &&  {\cal L}_{{\rm gal},1}(\phi) ={1\over 4H^2} \sqrt{-\gamma}\Phi^{\mu\nu}X^{(0)}\mn(\Phi) \, ,\nn\\
&&  {\cal L}_{{\rm gal},2}(\phi) ={1\over 6H^2} \sqrt{-\gamma}\Phi^{\mu\nu}X^{(1)}\mn(\Phi)\, , \nn\\
&&  {\cal L}_{{\rm gal},3}(\phi) ={1\over 6H^2} \sqrt{-\gamma}\Phi^{\mu\nu}X^{(2)}\mn(\Phi) \, ,\nn\\
&&  {\cal L}_{{\rm gal},4}(\phi) ={1\over 4H^2} \sqrt{-\gamma}\Phi^{\mu\nu}X^{(3)}\mn(\Phi) \, ,
\eea
where
\ba
\label{eq:phimn2}
\Phi\mn=\nabla_\mu \nabla_\nu \phi+H^2 \phi\gamma\mn\,.
\ea
The tensor $\Phi\mn$ and therefore all these expressions are invariant under a de Sitter version of the Galileon shift symmetry.  Let $Z^A(x)$, $A=0,1,2,3,4$ be an embedding of de Sitter into five dimensional Minkowski space.  Then the Galileon shift symmetry is given by
\be
\delta_B \phi(x)=B_A Z^A(x)\,,
\ee
where $B_A$ are constants parameterizing the five shift symmetries.

\section{\stu fields on AdS\label{adssappendix}}

The derivation of the \stu fields on a maximally symmetric background was provided in \cite{deRham:2012kf}. For convenience we shall reproduce the derivation here focusing on the AdS case. This can be derived from the dS case by substituting $H^2\to -L^{-2}$ throughout the analysis. The derivation of the covariant \stu formalism was also studied in \cite{Gao:2014ula,Gao:2015xwa}.   The strategy to derive the \stu fields in AdS is to base ourselves on those in flat space by embedding AdS$_4$ into a $(2+3)$-dimensional Riemann-flat spacetime (corresponding to five-dimensional Minkowski with two times), and then projecting the flat 5-dimensional \stu fields back onto AdS$_4$.

Let AdS$_4$ be embedded into a $(2+3)$-dimensional Minkowski space with cartesian coordinates $Z^A$ and metric
\ba
\d s^2=\eta_{MN}\d Z^M \d Z^N\, .
\ea
The four-dimensional AdS space is realized as the hypersurface defined by
\ba
\label{eq:surface}
\eta_{MN}Z^M Z^N=-L^{2}\,.
\ea
At this stage one can perform a change of coordinates $\{Z^M\} \to \{X^M\}=\{Y,x^\mu\}$, so as to foliate the five-dimensional spacetime into four-dimensional AdS slices. The hypersurface \eqref{eq:surface} is then located at $Y=0$ in the new coordinate system and the induced metric on that surface is AdS$_4$, with the metric denoted by $\gamma_{\mu \nu}\d x^\mu \d x^\nu$.
 The metric in these coordinates is
\ba
\d s^2=G_{MN}\d X^{M}\d X^N=e^{2Y/L}(-\d Y^2+\gamma_{\mu \nu} \d x^\mu \d x^\nu).\label{adsslicemetre}
\ea
We have the following relation between the cartesian coordinates and the AdS$_4$ foliated coordinates on the ambient flat space,
\ba
\eta_{AB}Z^A Z^B=-L^2 e^{2Y/L}\,.
\label{coordrelation1e}
\ea
Now consider the vector field $\bar\phi^A$ given in the cartesian coordinates by
\ba
\bar\phi^A=Z^A\,,
\ea
\ie perpendicular to the AdS$_4$ foliation.
In the AdS foliated coordinates, it is given by
\ba
\bar\phi^M=\left(L,0,0,0,0\right)\,,
\label{vecrelation1e}
\ea
where it manifestly points along the $Y$ direction, perpendicular to the AdS$_4$ slices.

We now introduce the five \stu scalar fields $\phi^A(Z)$.
They describe a diffeomorphism from the ambient Minkowski space to itself, with the property that it leaves the AdS$_4$ surface at $Y=0$ invariant,
\ba
\eta_{AB}\phi^A\phi^B=-L^2,\ \ \ {\rm when}\ Y=0\,.
\label{exreeqe}
\ea
This constraint \eqref{exreeqe} reduces the number of \stu fields down to four.
To see this explicitly, consider the identity diffeomorphism given by $\phi^A=\bar \phi^A$ and departures from it described by $V^A$,
\ba
\phi^A=\bar\phi^A-V^A\, .
\label{expandaarve}
\ea
Then in cartesian coordinates where $\bar\phi^A=Z^A$, the constraint \eqref{exreeqe} is
\ba
\eta_{AB}\phi^A\phi^BZ^2-2\eta_{AB}\bar\phi^A V^B+\eta_{AB}V^AV^B= -L^2\, \ \ \  {\rm when}\ Y=0\,.
\label{rhlheq1}
\ea
Expressing this constraint in the AdS slicing \eqref{adsslicemetre}, and splitting
 $V^M=\left\{V^Y,A^\mu\right\}$ we get
\ba
\eta_{AB}\phi^A\phi^B &=& -L^2e^{2Y/L}-2 G_{YY}\bar\phi^Y V^Y +G_{YY}(V^Y)^2+\gamma_{\mu\nu}A^\mu A^\nu \nn\\
&=& e^{2Y/L}\left(-L^2+  2LV^Y-(V^Y)^2+A^2  \right)\, ,
\ea
where $A^2\equiv \gamma_{\mu\nu}A^\mu A^\nu$.
Then the constraint \eqref{rhlheq1} evaluated at $Y=0$ gives
\ba
2LV^Y-\(V^Y\)^2+A^2=0\,,
\ea
which allows us to solve for $V^Y$ in terms of the independent \stu fields $A^\mu$,
\ba
V^Y=L\left(1-\sqrt{1+{1\over L^2}A^2}\right) \ \ \  {\rm when}\ Y=0\,. \label{Ysolutione}
\ea
The minus sign in front of the square root is chosen so that $V^Y=0$ when $A^\mu=0$.

In this language, we first introduce the \stu  through the ambient metric $\eta_{AB}\rightarrow \partial_A{ \phi^C} \partial_B \phi^D$, and then pull them back to the AdS surface, giving
\begin{equation}
\gamma_{\mu \nu}(x) \rightarrow \tilde{\gamma}_{\mu \nu}=\left.{\partial Z^A\over \partial x^\mu}{\partial Z^B\over \partial x^\nu}\left(\eta_{CD}\partial_A{ \phi^C} \partial_B \phi^D\right)\right|_{Y=0}\, .
\end{equation}
The `St\"uckelbergized' AdS metric $\tilde \gamma\mn$ can be expressed in cartesian coordinates using the split \eqref{expandaarve} prior to performing a  changed of coordinates into AdS slicing,
\ba
\tilde{\gamma}_{\mu \nu}&=&\left.{\partial Z^A\over \partial x^\mu}{\partial Z^B\over \partial x^\nu}\left(\eta_{AB}-\partial_AV_B-\partial_BV_A+\partial_AV^C\partial_BV_C \right)\right|_{Y=0} \nn\\
&=&\left.{\partial X^M\over \partial x^\mu}{\partial X^N\over \partial x^\nu}\left(G_{MN}-\nabla_MV_N-\nabla_NV_M+\nabla_MV^P\nabla_NV_P \right)\right|_{Y=0}\, . \label{xsolutionssse}
\ea
For convenience, we define the four-dimensional tensor
\ba
S_{\mu\nu}\equiv \left.{\partial X^M\over \partial x^\mu}{\partial X^N\over \partial x^\nu}\nabla_MV_N \right|_{Y=0}=\partial_\mu A_\nu-\Gamma^\lambda_{\mu\nu}V_\lambda-\Gamma^Y_{\mu\nu}V_Y\,,
\ea
where the Christoffel symbol $\Gamma^\lambda_{\mu\nu}$ on the right hand side is taken with respect to the four-dimensional AdS metric $\gamma\mn$, and $\Gamma^{Y}_{\mu\nu}={1\over L}\gamma_{\mu\nu}$. Then using \eqref{Ysolutione}, we have
\ba
S_{\mu\nu}=  \nabla_\mu A_\nu+\gamma_{\mu\nu}\left(1-\sqrt{1+{1\over L^2}A^2}\right) \,.
\ea

Using the expression for $V^Y$ given in \eqref{Ysolutione}, we have
\ba && \nabla_\mu V^Y =-{1\over L \sqrt{1+{1\over L^2}A^2} }T_{\mu}  \nn\\
&& {\rm where} \ \ T_{\mu}\equiv {1\over 2}\partial_\mu (A^2)-\sqrt{1+{1\over L^2}A^2}A_\mu \,.
\ea

Plugging these ingredients into the expression \eqref{xsolutionssse} for $\tilde \gamma\mn$, we obtain finally
\ba
\tilde{\gamma}_{\mu \nu}={\gamma}_{\mu \nu}-S_{\mu\nu}-S_{\nu\mu}+S_{\mu}^{\ \lambda}S_{\nu\lambda}-{1\over L^2+A^2}T_{\mu} T_{\nu}\,,
\ea
with
\ba
S_{\mu\nu}=  \nabla_\mu A_\nu+\gamma_{\mu\nu}\left(1-\sqrt{1+{1\over L^2}A^2}\right),\ \ \  T_{\mu}\equiv {1\over 2}\partial_\mu (A^2)-\sqrt{1+{1\over L^2}A^2}A_\mu.
\ea
Our \stu expressions \eqref{nonlineadsstue} and \eqref{nonlineadsstue2} are then obtained after appropriately canonically normalizing $A_\mu\rightarrow \bar A_\mu={2\over\mpl m} A_\mu$.

\bibliographystyle{utphys}
\addcontentsline{toc}{section}{References}
\bibliography{longdraft_arxiv_v2}

\end{document}